\documentclass[twocolumn]{aastex63}
\usepackage{multirow}

\begin{document}
\title{Multiwavelength Characterization of the High Mass X-ray Binary Population of M33}
\author[0000-0003-3252-352X]{Margaret Lazzarini}
\affiliation{Division of Physics, Mathematics and Astronomy, California Institute of Technology, 1200 E California Blvd., Pasadena, CA 91125, USA}
\author[0000-0002-0206-1208]{Kyros Hinton}
\affiliation{Department of Astronomy, Box 351580, University of Washington, Seattle, WA 98195, USA}
\author[0000-0003-1247-9349]{Cheyanne Shariat}
\affiliation{Department of Physics and Astronomy, University of California Los Angeles, Los Angeles, 430 Portola Plaza, Los Angeles, CA 90095, USA}
\author[0000-0002-7502-0597]{Benjamin F. Williams}
\affiliation{Department of Astronomy, Box 351580, University of Washington, Seattle, WA 98195, USA}
\author[0000-0002-9202-8689]{Kristen Garofali}
\affiliation{NASA Goddard Space Flight Center, Code 662, Greenbelt, MD 20771, USA}
\author[0000-0002-1264-2006]{Julianne J. Dalcanton}
\affiliation{Center for Computational Astrophysics, Flatiron Institute, 162 Fifth Avenue, New York, NY 10010, USA}
\affiliation{Department of Astronomy, Box 351580, University of Washington, Seattle, WA 98195, USA}
\author[0000-0001-7531-9815]{Meredith Durbin}
\affiliation{Department of Astronomy, Box 351580, University of Washington, Seattle, WA 98195, USA}
\author[0000-0001-7539-1593]{Vallia Antoniou}
\affiliation{Department of Physics \& Astronomy, Texas Tech University, Lubbock, TX 79409, USA}
\affiliation{Center for Astrophysics | Harvard \& Smithsonian, 60 Garden St., Cambridge, MA 02138, USA}
\author[0000-0002-4955-0471]{Breanna Binder}
\affiliation{Department of Physics \& Astronomy, California State Polytechnic University, 3801 W. Temple Ave, Pomona, CA 91768, USA}
\author[0000-0002-3719-940X]{Michael Eracleous}
\affiliation{Department of Astronomy \& Astrophysics and Institute for Gravitation and the Cosmos, The Pennsylvania State University, 525 Davey Lab, University Par, PA 16802}
\author[0000-0001-7855-8336]{Neven Vulic}
\affiliation{Eureka Scientific, Inc., 2452 Delmer St., Suite 100, Oakland, CA 94602-3017, USA}
\affiliation{NASA Goddard Space Flight Center, Code 662, Greenbelt, MD 20771, USA}
\affiliation{Department of Astronomy, University of Maryland, College Park, MD 20742-2421, USA}
\affiliation{Center for Research and Exploration in Space Science and Technology, NASA/GSFC, Greenbelt, MD 20771, USA}
\author[0000-0002-7584-4756]{Jun Yang}
\affiliation{Department of Physics and Kavli Institute for Astrophysics and Space Research, Massachusetts Institute of Technology, Cambridge, MA 02139, USA}
\author[0000-0001-9110-2245]{Daniel Wik}
\affiliation{University of Utah, Salt Lake City, UT 84112, USA}
\author{Aria Gasca}
\affiliation{PUC Early College Academy for Leaders and Scholars, 2050 N San Fernando Rd, Los Angeles, CA 90065, USA}
\author{Quetzalcoatl Kuauhtzin}
\affiliation{PUC Early College Academy for Leaders and Scholars, 2050 N San Fernando Rd, Los Angeles, CA 90065, USA}

\correspondingauthor{Margaret Lazzarini}
\email{mlazz@caltech.edu}

\begin{abstract}
We present multi-wavelength characterization of 65 high mass X-ray binary (HMXB) candidates in M33. We use the Chandra ACIS survey of M33 (ChASeM33) catalog to select hard X-ray point sources that are spatially coincident with UV-bright point source optical counterparts in the Panchromatic Hubble Andromeda Treasury: Triangulum Extended Region (PHATTER) catalog, which covers the inner disk of M33 at near infrared, optical, and near ultraviolet wavelengths. We perform spectral energy distribution (SED) fitting on multi-band photometry for each point source optical counterpart to measure its physical properties including mass, temperature, luminosity, and radius. We find that the majority of the HMXB companion star candidates are likely B-type main sequence stars, suggesting that the HMXB population of M33 is dominated by Be-XRBs, as is seen in other Local Group galaxies. We use spatially-resolved recent star formation history (SFH) maps of M33 to measure the age distribution of the HMXB candidate sample and the HMXB production rate for M33. We find a bimodal distribution for the HMXB production rate over the last 80 Myr, with a peak at $\sim$10 Myr and $\sim$40 Myr, which match theoretical formation timescales for the most massive HMXBs and Be X-ray binaries (Be-XRBs), respectively. We measure an HMXB production rate of 107$-$136 HMXBs/(M$_{\odot}$ yr$^{-1}$) over the last 50 Myr and 150$-$199 HMXBs/(M$_{\odot}$ yr$^{-1}$) over the last 80 Myr. For sources with compact object classifications from overlapping NuSTAR observations, we find a preference for giant/supergiant companion stars in BH-HMXBs and main sequence companion stars in neutron star HMXBs (NS-HMXBs). 
\end{abstract}

\section{Introduction}
High mass X-ray binaries (HMXBs) are systems where a black hole or neutron star accretes material from its massive stellar companion. HMXBs are an important tool for studying accreting compact objects and massive binary stellar evolution. Nearby galaxies, including M33, provide an opportunity to study a population of HMXBs, allowing us to measure population demographics that can constrain theoretical models of massive binary stellar evolution.

The properties of HMXB populations scale with galaxy-wide properties such as star formation rate \citep[SFR;][]{Ranalli2003,Gilfanov2004,Antoniou2010,Mineo2012,Antoniou&Zezas2016,Lehmer2019,Antoniou2019}, metallicity \citep{Basuzych2013,BasuZych2016,Brorby2016}, stellar mass \citep{Lehmer2010,Antoniou2019}, and the number of O- and B-type stars \citep{Antoniou2019}. In nearby galaxies, we can place HMXBs in context of both their host galaxy and their more local parent stellar population. 

Massive stars ($\gtrsim$8 M$_{\odot}$) play a critical role in many astrophysical processes. During their lives they inject ionizing radiation into the interstellar medium (ISM) and at the end of their lives impact the evolution of galaxies via supernova feedback \citep[e.g.,][]{Dalgarno1972, Oey1999,Telford2023}. Most massive stars form in binary or higher order multiple systems and a majority of these systems have separations small enough for the stars to interact at some point during their lives \citep[e.g.,][]{Sana2012,Moe&DiStefano2017}. These interactions, including mass transfer and common envelopes, impact the evolution of the stars in the binary. Some massive stellar binaries go on to form binary compact objects which can merge, depending on their separation, within a Hubble time generating detectable gravitational waves \citep[e.g.,][]{Tauris2017}. HMXBs are also thought to be a major contributor of ionizing radiation in the early universe, likely playing a key role in cosmic reionization \citep[e.g.,][]{Justham&Schawinski2012,Mesinger,Madau2017,Greig2018}.

Given their importance in many fields of astrophysics, HMXBs have been studied extensively in the Milky Way and other galaxies. Populations of HMXBs can be studied in nearby galaxies with relatively few telescope pointings, and the proximity of these galaxies allows us to resolve their individual stars, which is necessary for identifying the companion star to the accreting compact object in the HMXB system. The populations of HMXBs in the Large and Small Magellanic Clouds have been studied extensively at both X-ray and optical wavelengths. These studies have identified the companion star spectral type and compact object type for over 200 HMXBs in both galaxies combined, providing an insight to HMXB formation at low metallicity and high SFR \citep[e.g.,][]{Liu2006,Antoniou2009,Antoniou2010,Antoniou&Zezas2016,yang2017comprehensive,Haberl2016,Antoniou2019,Haberl2022}. In M31, overlapping Hubble Space Telescope (HST) and Chandra observations have been used to measure HMXB population demographics in a large spiral galaxy of similar mass and metallicity to the Milky Way \citep{Williams2018,Lazzarini2018,Lazzarini2021}. 

M33 is an obvious site to study HMXBs in the Local Group. It is the third largest spiral galaxy in the Local Group and has a measured metallicity gradient; the center of the galaxy is slightly super-solar and the metallicity drops to roughly that of the LMC at the edges of the disk \citep{carrera2008,Cioni2009,Magrini2009,Magrini2010,beasley2015,ToribioSanCipriano2016,Lin2017}. Although it is about ten times less massive than M31 or the Milky Way \citep[e.g.,][]{Quirk2022}, it has a higher SFR intensity \citep[SFR per area;][]{Verley2009,Lewis}, making it an excellent site to study young stellar systems. Because of its high SFR, M33 also hosts a large population of supernova remnants (SNRs), which have been studied at X-ray and optical wavelengths \citep[e.g.,][]{Long2010,Garofali2017,Koplitz2023}.

M33 has been observed extensively at X-ray wavelengths with the Einstein Observatory \citep{Long1981,Markert&Rallis1983,Trinchieri1988}, ROSAT \citep{Schulman1995,Long1996,Haberl&Pietsch2001}, XMM-Newton \citep{Pietsch2004,Misanovic2006,Williams2015}, Chandra \citep{Grimm2005,Tullmann2011}, and NuSTAR \citep{Yang2022}. The Chandra ACIS Survey of M33 (ChASeM33) observed 70\% of the area within the D$_{25}$ isophote of M33, providing a deep catalog of X-ray point sources with high angular resolution, which is critical for identifying the optical counterparts to accreting compact objects in HMXBs. A uniform survey of M33 at optical wavelengths was not available until recently with the Panchromatic Hubble Andromeda Treasury: Triangulum Extended Region (PHATTER), an HST survey of the inner disk of M33 in six photometric bands spanning near-infrared through near-ultraviolet wavelengths \citep{Williams2021}. The survey's photometry catalog contains six-band photometry for over 20 million individual stars and has been used to study the galaxy's recent star formation history \citep[SFH;][]{Lazzarini2022}, structure (A. Smercina et al. 2023, in prep.), and population of star clusters \citep{Wainer2022,Johnson2022}. 

Previous studies of the HMXB population of M33 have led to both detailed characterization of a few individual sources and properties of the population. \citet{Garofali2018} performed a population study using archival HST observations and found a peak in the M33 HMXB age distribution at $<$5 Myr and another peak at $\sim$40 Myr, with a valley between these two peaks with little to no HMXB production. Previous studies of HMXBs in M33 focused on individual systems. The nucleus of M33 hosts the nearest ultraluminous X-ray source (ULX) \citep{Long2002}, X-8. The X-ray emission is powered by super-Eddington accretion, although whether the compact object is a black hole or neutron star has not yet been firmly determined \citep{Krivonos2018,West2018}. There are also two eclipsing X-ray binaries in M33 \citep{Pietsch2004,Pietsch2006,Pietsch2009}. One of the two, M33 X-7, comprises a massive stellar-mass black hole in orbit with an O-type stellar companion \citep{Orosz2007,Ramachandran2022}. The exact orbital separation and masses of the black hole and companion star in M33 X-7 have been debated, but the system has provided interesting constraints for massive binary stellar evolution models. The other eclipsing binary in M33 \citep{Pietsch2004} has been less well studied and is not covered in the footprint of our analysis in this paper.

In this paper, we leverage the deep, uniform coverage of the inner disk of M33 at near-infrared/optical/near-ultraviolet wavelengths to systematically identify HMXB companion star candidates. We perform spectral energy distribution (SED) fitting on the HMXB companion star candidates to infer their physical properties including temperature, radius, luminosity, and mass. We use the spatially-resolved recent SFH of M33, measured using the optical PHATTER photometric catalog, to place constraints on the age distribution of the HMXB population of M33 and its HMXB production rate over time. 

We describe the X-ray and near-infrared/optical/near-ultraviolet observations used in this analysis in Section \ref{sec:data}. In Section \ref{sec:analysis} we describe how we identify and evaluate the quality of the HMXB candidate sample and our methodology for SED-fitting of the companion star candidates and measurement of the HMXB candidate ages using the SFH maps. We include a discussion of our results in Section \ref{sec:results} and summarize our findings in Section \ref{sec:conclusion}. We assume a distance to M33 of 859 kpc, or a distance modulus of 24.67, throughout our analysis \citep{Grijs2014}.
 
\section{Data}\label{sec:data}
\begin{figure*}
\centering
\includegraphics[width=0.99\textwidth]{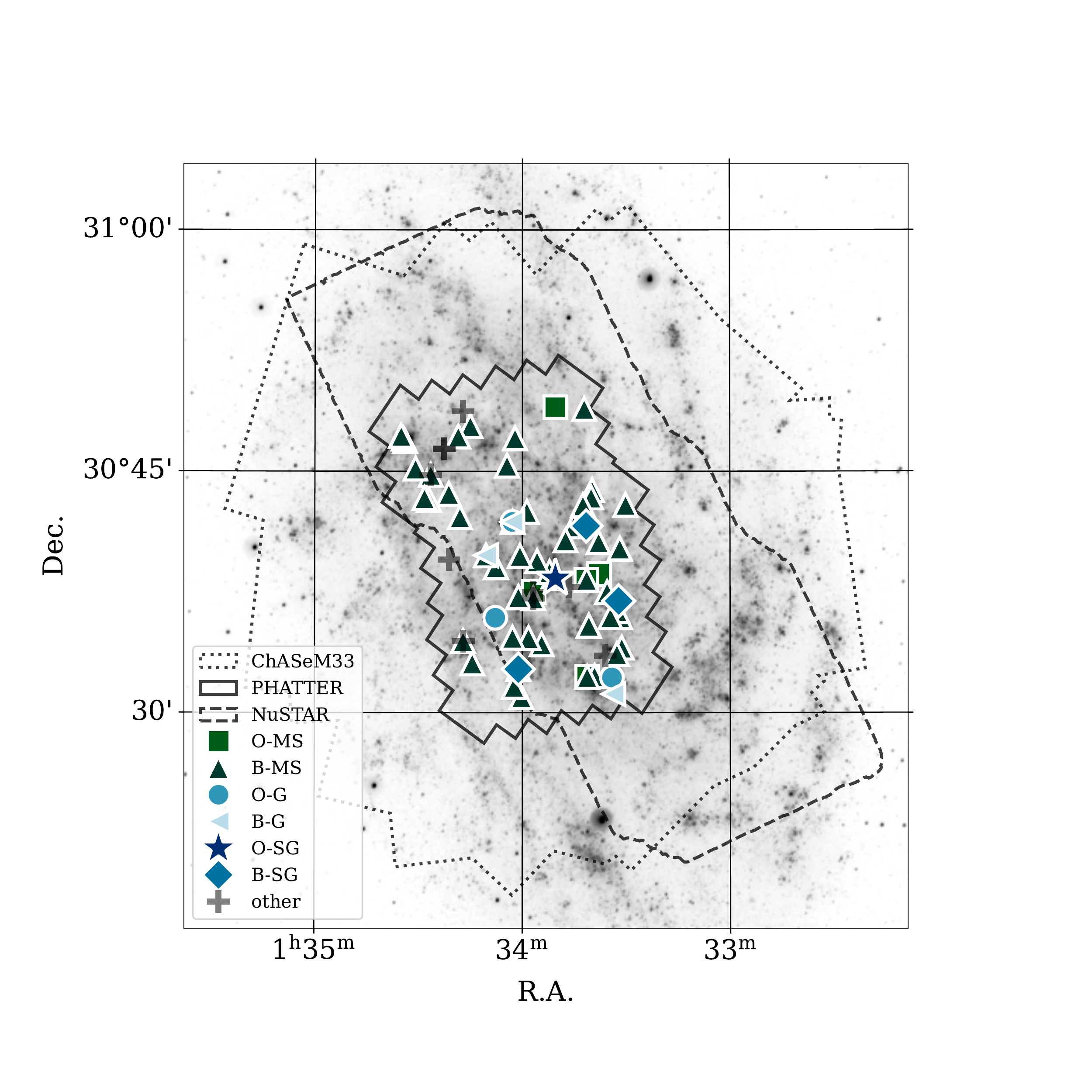}
\caption{Near-UV image of M33 with the positions of HMXB companion star candidates plotted. The shape and color of each point indicates its most likely spectral type, O- or B-type main sequence (MS), giant (G) or supergiant (SG), described in more detail in Section \ref{sec:SED-fitting}. The outline of three multi-wavelength surveys of M33 used in our analysis are overplotted. The footprint of the Chandra survey used in this analysis \citep[ChASeM33;][]{Tullmann2011}, is outlined with a dotted black line. The HST survey we use to identify HMXB companion star candidates, is outlined in the solid black line \citep[PHATTER;][]{Williams2021}. We also outline the overlapping NuSTAR survey of M33 \citep{Yang2022}.}
\label{fig:M33_overview}
\end{figure*}

We use X-ray and optical/near IR/near UV imaging and photometric catalogs of the inner disk of M33 to identify our HMXB candidate sample. We use the final source catalog from the ChASeM33 survey \citep{Tullmann2011} and the optical/near IR/near UV catalog from the PHATTER survey \citep{Williams2021}. In Figure \ref{fig:M33_overview} we show the outline of each survey. We provide more details on each survey below.

\subsection{X-ray Data}
The ChASeM33 ACIS survey covered $\sim$70\% of the D$_{25}$ area of M33 in the 0.35$-$8.0 keV band \citep{Tullmann2011}. The resulting catalog includes 662 sources, 202 of which lie within the PHATTER survey footprint in the inner disk of M33. The catalog has a limiting unabsorbed luminosity of $\sim2.4 \times 10^{34}$ erg s$^{-1}$ in the full 0.35$-$8.0 keV band and includes positions, fluxes, and net counts for each source in multiple energy bands.

We use the X-ray source positions from this catalog, along with their associated errors, to identify the optical counterpart candidates for each X-ray source, thus creating our sample of HMXB candidates. We also use count rates and fluxes, and spectral fits for sources where these are available, from the ChASeM33 catalog when evaluating the quality of our HMXB sample. We provide more details on how we identify optical counterparts to X-ray sources in Section \ref{sec:counterpart_id} and how we evaluate the quality of our HMXB candidates in Section \ref{sec:flags}.

In Table \ref{table:beast_table} we include a list of HMXB candidates. We include each source's identification number, X-ray source position, and 0.35$-$8.0 keV luminosity from \citep{Tullmann2011}. We calculated the luminosity using the count rates published by \citet{Tullmann2011}. For this calculation we assume a power law index of 1.7 and a foreground N$_{H}$ of 5$\times$10$^{20}$ cm$^{-2}$ \citep{Lebouteiller2006}. We used the Chandra proposal planning toolkit PIMMS tool to derive a conversion factor of 1.087$\times 10^{-11}$ from count rate (counts s$^{-1}$) to flux , which we then convert to luminosity.

We also cross-matched each X-ray source with other X-ray surveys of M33 including \citet[][Chandra]{Grimm2005}, \citet[][XMM-Newton]{Williams2015}, and \citet[][NuSTAR]{Yang2022}. We include the counterpart of each source in the aforementioned catalogs in Table \ref{table:beast_table} and include their classifications (if provided) in each catalog. We describe our cross-matching methodology in Section \ref{sec:cross-match}. 

\subsection{Optical Data}
We use the photometric catalog and imaging from the PHATTER survey in our data analysis \citep{Williams2021}. The PHATTER survey covered an area of 14 kpc$^{2}$ in the inner disk of M33, providing six-band photometry for over 22 million stars. In this study, we use the mosaic imaging from the PHATTER survey and the photometric catalog \citep{Williams2021} to identify the optical counterpart for \citet{Tullmann2011} X-ray sources within the PHATTER survey footprint. We also use the spatially resolved recent SFH maps derived from the PHATTER optical photometry by \citet{Lazzarini2022} to measure the age distribution and HMXB production rate for M33.

\subsection{Astrometric Alignment}
We use the X-ray source properties and positions from the ChASeM33 survey \citep{Tullmann2011} and optical imaging and photometric catalogs from PHATTER \citep{Williams2021} in our analysis. The ChASeM33 catalog is aligned to 2MASS \citep{2MASS_Catalog,Tullmann2011,Garofali2018} and the PHATTER catalog is aligned to GAIA. To ensure that our X-ray and optical data were aligned, we selected sources in the PHATTER catalog with a F160W magnitude $<$15.5, to match the 99.9\% completeness limit of the 2MASS survey in the H band, which closely overlaps with the F160W filter on HST. We then matched the 2MASS catalog to the PHATTER sources. We performed this task in an iterative fashion so that false matches could be culled from our list of matches.  

To confirm our matches, we plotted the PHATTER photometry and the 2MASS photometry for each source to create a spectral energy distribution (SED). We visually inspected the SED for each match to confirm that the SED shape looked smooth at the transition from the HST to 2MASS filters. We also plotted the positions of all stars within 5 arcsec that fit our PHATTER criteria (F160W$<$15.5) and all 2MASS stars that lie within the same area to confirm that the sources look like a good match. Due to our magnitude cuts, the source density of PHATTER sources is greatly reduced, reducing the chances of an erroneous match. 

We find an offset between the two catalogs of $\Delta$R.A. 0.04 arcsec and $\Delta$Dec. 0.004 arcsec, with a RMS of 0.1 in both R.A. and Dec. The offset in both R.A. and Dec. is of the same order of magnitude as the mean 2MASS positional errors for the sources we use as matches in our alignment. Because the offset between the two catalogs is less than the mean 2MASS positional error for all sources, we did not update the positions of the sources in the ChASeM33 catalog (aligned the 2MASS) or the PHATTER catalog.

\section{Analysis}\label{sec:analysis}

We use near-UV/optical and X-ray observations to identify 65 HMXB candidates in M33. We perform SED fitting to determine the likely physical properties of candidate HMXB companion stars. Using maps of the spatially resolved recent SFH of M33, we measure the age distribution and production rate for the HMXB candidate sample. We use multi-wavelength information including Chandra hardness ratios, UV/optical SED shapes, local star formation rates, and measurements of extinction to evaluate our HMXB candidate sample and identify the highest quality HMXB candidates. Lastly, we cross-match our HMXB candidate sample with previous catalogs of X-ray sources and HMXB candidates in M33.

\subsection{Identification of HMXB Candidate Sample}\label{sec:counterpart_id}
We identify HMXB candidates using HST optical/near-UV imaging and photometry from the PHATTER survey \citep{Williams2021} and Chandra-detected X-ray source positions and fluxes/count rates from the ChASeM33 survey \citep{Tullmann2011}. As described in more detail in this section, we identify 64 HMXB candidates using the methodology described here. We include M33 X-8, the nuclear ULX, in our sample when analyzing the HMXB ages and production rates, which brings our total HMXB candidate sample of 65 sources.

To identify our sample of HMXB candidates, we started with all X-ray point sources from \citet{Tullmann2011} that were within the PHATTER survey footprint that were not classified as AGN based on optical spectroscopy \citep{Tullmann2011}, supernova remnants based on their X-ray, radio, and optical properties \citep{Long2010}, or colliding wind binaries \citep{Garofali2019}. We also removed X-ray sources from our sample if the optical counterpart identified in the PHATTER imaging was a resolved galaxy. We include the nuclear HMXB (X-8; ChASeM33 318) in our HMXB candidate sample for age and production rate measurements, although we cannot resolve its optical counterpart within the nuclear star cluster of M33.

We selected optical counterpart candidates from within 1.5$\sigma$ of the X-ray source position, using the errors and X-ray source positions presented in the ChASeM33 survey catalog. This radius was chosen to maximize the recovery of the HMXB population in M33, while minimizing chance superpositions between X-ray sources and optical counterpart candidates. To settle on this distance, we started with a 1$\sigma$ error circle and expanded the radius iteratively by 0.1$\sigma$. We expanded out to 1.5$\sigma$, which allowed us to recover the HMXB candidates identified by \citet{Garofali2018} that fall within the PHATTER footprint but that were not otherwise disqualified (i.e. likely SNR, resolved galaxy in PHATTER imaging, colliding-wind binary).

When only selecting companion star candidates from within 1$\sigma$, we identified 40 unique HMXB candidates, and when we select from within 1.5$\sigma$ we identify 64 unique HMXB candidates. As described in more detail in Section \ref{chance_superposition}, with a match radius of 1.5$\sigma$ we expect 3$-$4 false HMXBs due to a chance superposition of a star in the PHATTER catalog that meets our selection criteria and an X-ray source in the ChASeM33 source catalog. With a match radius of 1$\sigma$ we expect 1$-$2 false HMXBs due to a chance superposition. When we expand our match radius from 1$\sigma$ to 1.5$\sigma$ we gain 24 HMXB candidates, suggesting that the majority of the gained HMXB candidates are not false matches. We expect the X-ray source to be located within the 1$\sigma$ error on its positions 68\% of the time, and when we expand to 1.5$\sigma$ we expect the X-ray source to be located within this 1.5$\sigma$ positional error 87\% of the time. We find that the trade off of a slight increase in the number of chance superpositions is outweighed by the increase in our sample of HXMB candidates.

We selected optical counterpart candidates that had colors and magnitudes in the PHATTER photometric catalog consistent with being massive (M$\gtrsim$8 M$_{\odot}$) main sequence, giant, or supergiant stars. We chose this mass cut off because it is the known lower limit for the masses of Be stars in Be-XRBs \citep{Shao&Li2014,Coe&Kirk2015}. We used Padova stellar models \citep{Marigo2008} to define the ranges of colors and magnitudes of massive main sequence and giant/supergiant stars. To define the ranges in color and magnitude, we started with Padova stellar tracks and plotted them onto optical and UV color-magnitude diagrams (CMDs). We then applied extinction to each photometric band for our model stars using the coefficients for reddening in the HST bandpasses used in the PHATTER survey \citep{Schlafly&Finkbeiner2011} and applied 1 and 2 magnitudes of dust extinction (A$_{V}$). To ensure that we selected a complete sample, we included stars within 0.5 mag of the lowest expected magnitude for an 8 M$_{\odot}$ star. This margin makes us potentially sensitive to unreddened counterpart candidates down to $\sim$6 M$_{\odot}$.

We used four sets of criteria to select likely massive stars associated with X-ray point sources. We list the exact color and magnitude cuts we used to select HMXB companion star candidates in Table \ref{table:comp_star_criteria}. Counterparts were selected if they were within 1.5$\sigma$ of an X-ray point source position and had colors consistent with being a massive (M$\gtrsim$8 M$_{\odot}$) main sequence or giant/supergiant star with up to two magnitudes of dust extinction, A$_{V}$. 

Using these criteria, we identify 64 HMXB candidates. The 65th HMXB we include in our sample is the ULX in the nuclear cluster of M33, X-8 (ChASeM33 318), which does not have an identified companion star candidate due to high stellar density in the central region of the galaxy. In our identified HMXB candidate sample, 23 X-ray sources have more than one star that meets one of the aforementioned criteria. We list the HST photometry for all potential HMXB companion star candidates in Table \ref{table:phatter_photometry} and list the PHATTER position and which ChASeM33 source the star is associated with in Table \ref{table:beast_table}.

\begin{figure*}
\centering
\includegraphics[width=0.99\textwidth]{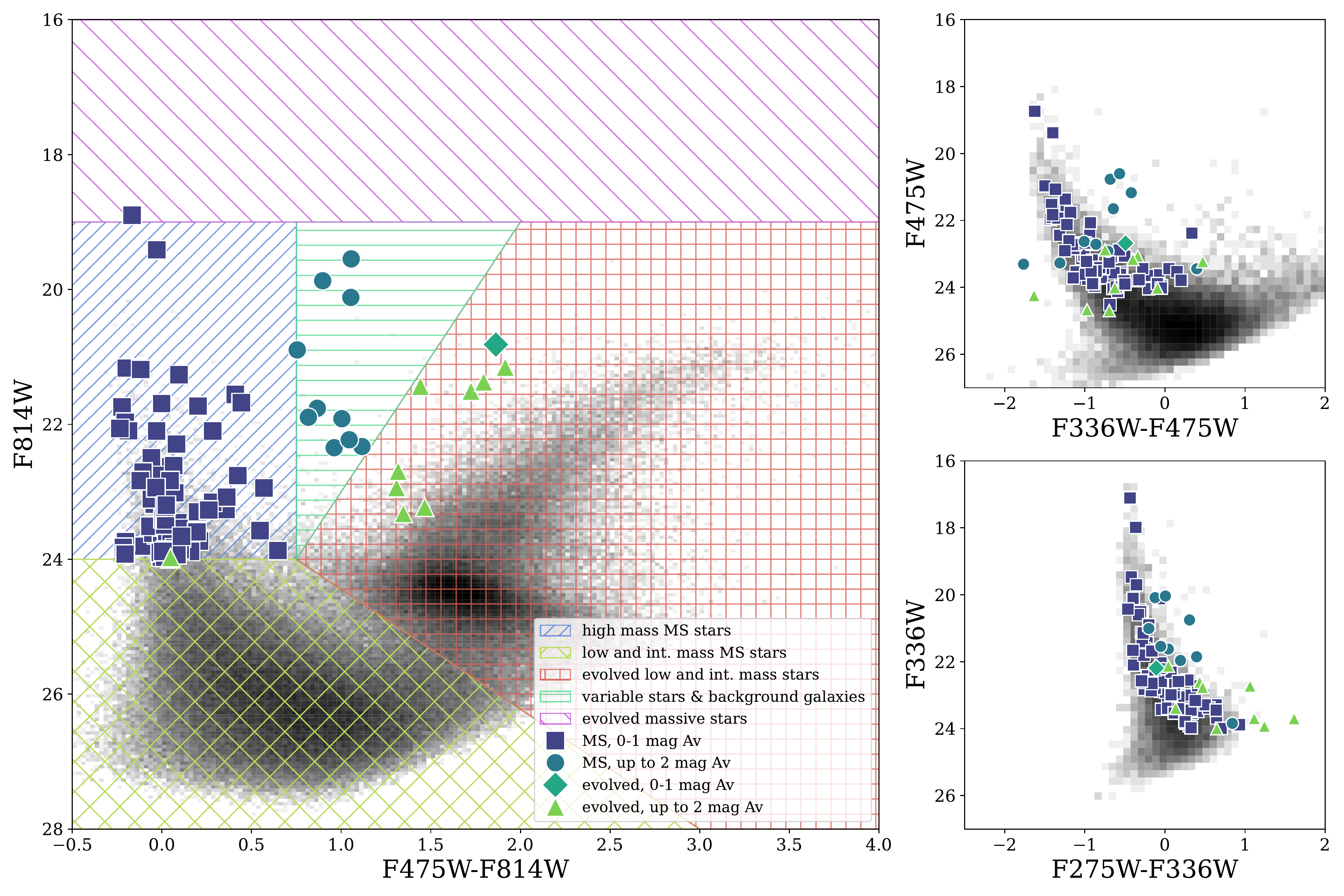}
\caption{Optical and ultraviolet CMDs showing HMXB companion star candidates color-coded for the criteria used to select it as such, outlined in the legend and described in detail in Section \ref{sec:counterpart_id}. \textbf{left:} Optical CMD, the background grayscale histogram represent all stars in the PHATTER photometric catalog within 5 arcseconds of an X-ray source in the ChASeM33 catalog, used here to outline the main features of the CMD. The shaded, colored regions show the areas of the CMD occupied by stars of different masses and evolutionary stages. \textbf{upper right:} Optical-ultraviolet CMD. Only the main sequence shows up on this CMD because older, redder stars do not have measured ultraviolet (F336W) magnitudes. The HMXB companion star candidates all lie on or just red-ward of the main sequence on this CMD. \textbf{lower right:} UV-only CMD}
\label{fig:HMXB_CMDs}
\end{figure*}

In Figure \ref{fig:HMXB_CMDs} we present color-magnitude diagrams showing the location of the optical counterparts we identified using the criteria described above. While the location of the HMXB companion star candidates spans about 2 magnitudes in F475W$-$F814W color (left CMD) we see that all of these stars lie on or close to the main sequence in the optical-UV (upper right) and UV-only (lower right) CMDs, suggesting these stars are likely to be massive stars.

\subsubsection{Chance Superposition}\label{chance_superposition}
We calculated the number of spurious HMXB companion star candidates we expect based on a chance superposition of an OB star within the 1.5$\sigma$ error circle of our X-ray sources. We applied the selection criteria outlined above to the full PHATTER photometric catalog to find the total of stars in the PHATTER catalog that met our criteria, and turned this into a density of stars (arcsec$^{-2}$) when divided by the total area of the PHATTER survey. We then summed the area of the 1.5$\sigma$ error circles for all ChASeM33 X-ray sources (regardless of optical counterpart type) that fall within the PHATTER survey footprint. We multiplied this area by the density of OB stars that meet our selection criteria in the PHATTER catalog to find the total number of chance superpositions we expect. We find that there should be 3$-$4 chance superpositions, so we expect that $<$10\% of our HMXB candidates could be spurious. 

\subsubsection{AGN Contamination}
In the process of identifying HMXB companion star candidates, we visually inspected the PHATTER imaging at the location of each ChASeM33 X-ray source within the PHATTER survey footprint. We identified 29 resolved background galaxies that were spatially associated with ChASeM33 X-ray sources. 

There are 202 ChASeM33 X-ray sources that lie within the footprint of the PHATTER survey.
Using the logN$-$logS relation presented in \citet{Tullmann2011} for AGN, we expect $\sim$100 AGN within the area of M33 that is covered by both the ChASeM33 and PHATTER surveys. There are 28 X-ray sources identified as SNRs by \citet{Long2010} using narrow band imaging. We identify 65 HMXB candidates using the methods described in Section \ref{sec:counterpart_id}. 

We note that while we expect there to be $\sim$100 background AGN in the area of the PHATTER survey in M33, we only identify 29 through visual inspection of the PHATTER HST images. This difference is likely due to the limited depth of the PHATTER imaging and difficulty detecting and resolving the disk of a galaxy seen through the disk of M33, particularly in regions with high stellar density and/or dust extinction. We also expect our quality cuts to remove background AGN with point source optical counterparts from our sample, as the optical/UV SEDs for these sources should differ in shape from massive stars in the disk of M33. We describe this series of quality cuts that we perform on our sample in more detail in Section \ref{sec:flags}.

\subsection{SED Fitting}\label{sec:SED-fitting}
We  used the Bayesian Extinction and Stellar Tool \citep[BEAST;][]{BEAST} software to fit SEDs for the identified point source optical counterparts. The BEAST fits observed SEDs with theoretical SEDs from the Padova/PARSEC single-star stellar evolution models \citep{Marigo2008,bressan2012,marigo2017} using Bayesian methods \citep{BEAST}. We use the four optical to near-ultraviolet bands from the PHATTER survey in our fitting: F275W, F336W, F475W, F814W bands, with central wavelengths 2750, 3375, 4750, and 8353 \AA$ $, respectively. When performing our SED fits, we assume that all sources are in the disk of M33, at a fixed distance modulus of 24.67, or 859 kpc \citep{Grijs2014}, and are moving with a constant radial velocity of -179.2 km s$^{-1}$ \citep{McConnachie2012}. 

The BEAST fits for parameters including age, mass, metallicity, distance, dust column density (A$_{V}$), average grain size (R$_{V}$), and f$_{A}$---a parameter describing the distribution of different types of dust observed in the Local Group. From these parameters, the BEAST derives the luminosity, effective temperature, radius, and surface gravity for each HMXB companion candidate. 

The BEAST imposes a Kroupa initial mass function \citep{Kroupa2001} as a prior on stellar mass, a uniform prior for A$_{V}$, R$_{V}$, f$_{A}$, stellar age, and stellar metallicity \citep{BEAST}. The BEAST then maps the initial mass and log(t) onto  log(T$_{eff}$) vs. log(L) and log(T$_{eff}$) vs. log(g) diagrams to produce priors on the other stellar physical parameters–including luminosity, effective temperature, radius, and surface gravity–which reflect the expected distribution of known stars \citep{BEAST}. The BEAST returns the parameters for the best-fit stellar and dust model and probability distributions for each parameter. We use these probability distributions to report the 16th and 84th percentile errors for each parameter.

We allow the log(t) to range from 6 to 8 (10$^{6}$ -- 10$^{8} yr$) in steps of 0.1. The metallicity grid was constrained to $Z_{initial}$ = [0.03, 0.019, 0.012, 0.008, 0.004, 0.001,0.0004], which covers the slightly supersolar to solar range as seen in M33 \citep{Cioni2009}. We allow A$_V$ values ranging from 0.0 to 5.0 mag in steps of 0.25 mag. We chose this range to be consistent with the values of A$_{V}$ measured by \citet{Lazzarini2022} when recovering the spatially resolved recent SFH of M33, but allowing for higher A$_{V}$ to account for the fact that HMXBs may be in areas of their local stellar environment with higher extinction than average.

\begin{figure*}
\centering
\includegraphics[width=0.85\textwidth]{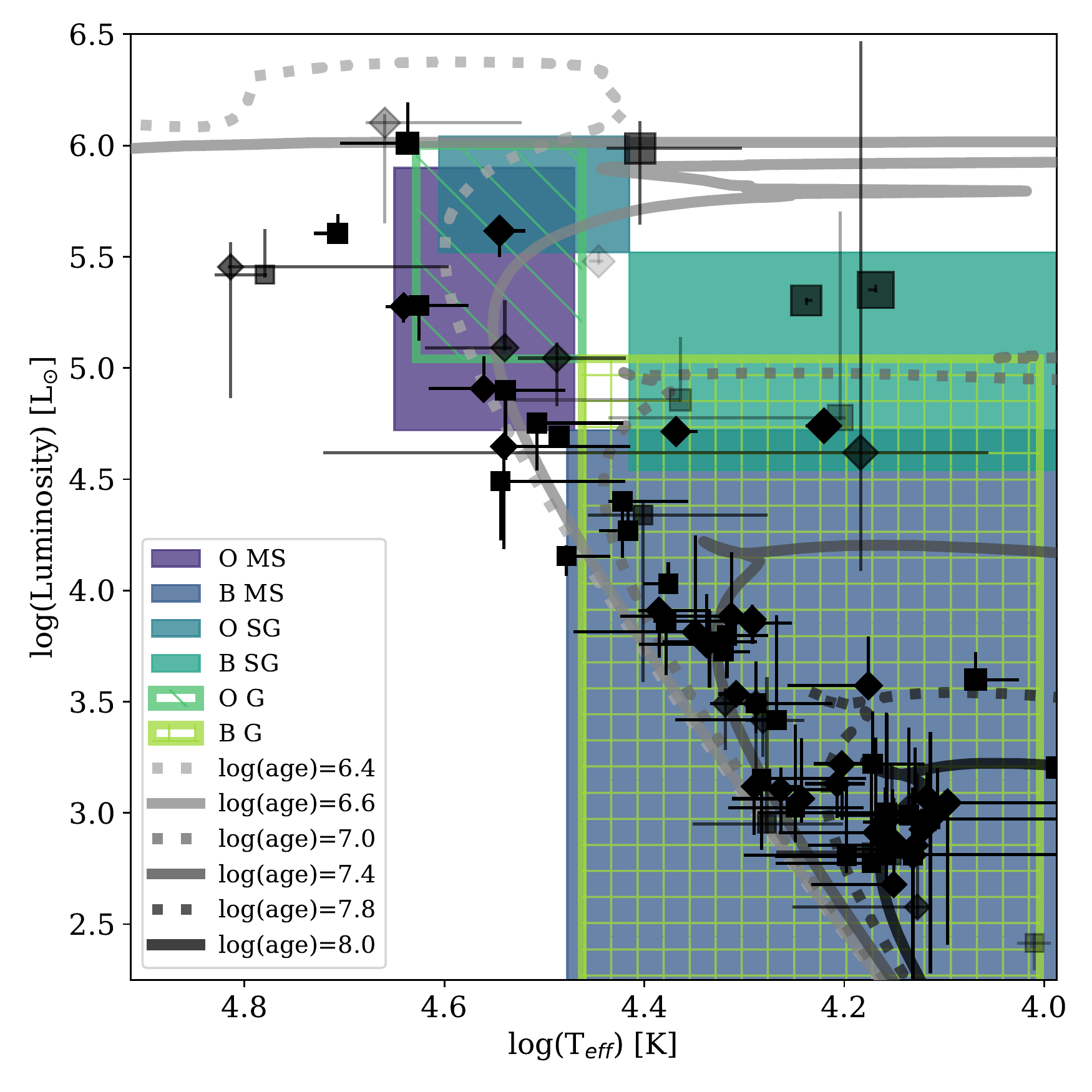}
\caption{Hertzsprung-Russell diagram with the SED-fit derived effective temperatures and luminosities for our HMXB companion star candidates. Each HMXB companion star candidate is plotted as an individual point with errors. The shape of the point denotes the number of companion star candidates, diamond-shaped points represent HMXBs with only one companion star candidate and square-shaped points represent HMXBs with multiple companion star candidates. The size of the point scales with the best-fit radius of the source. The opacity of the point scales with its ``quality'' according to the evaluation criteria described in Section \ref{sec:flags}. The background shaded regions indicate the expected ranges in log(T$_{eff}$) and log(Luminosity) for O and B type main sequence (MS), giant (G), and supergiant (SG) stars from \citet{Lamers2017}, also listed in Table \ref{table:sp_type_ranges}. Isochrones from the Padova stellar models \citep{Marigo2008} are over-plotted.}
\label{fig:HR_diagram}
\end{figure*}

We determine the most likely spectral type for each source based on its best-fit effective temperature, luminosity and stellar radius. In Figure \ref{fig:HR_diagram}, we plot the Hertzsprung-Russell diagram for all HMXB companion star candidates using the best-fit luminosity and effective temperature. For HMXBs with only one companion star candidate, we plot the star with a diamond-shaped point. For HMXBs with multiple companion star candidates, we plot each star with a square shaped point. The error bars represent the 16th and 84th percentile values corresponding to the $ \pm 1 \sigma$. If an error bar only extends in one direction, for example if a point does not have a lower error in log(T$_{eff}$), the best-fit value for the given parameter lies outside the 16th$-$84th percentile range. The size of each point scales with its best fit stellar radius. We also include background patches to show the expected ranges in effective temperature and luminosity for O- and B-type main sequence, giant, and supergiant stars \citep{Lamers2017}. We include isochrones from the Padova stellar models at different ages \citep{Marigo2008}.  We list the ranges of effective temperature, radius, and luminosity for each type of star in Table \ref{table:sp_type_ranges}.

\subsection{Age Determination with Spatially Resolved SFHs}\label{sec:age_determination}
We use the spatially resolved recent SFH of M33 derived from the PHATTER optical-only photometry catalog \citep{Lazzarini2022} to determine the age distribution and production rate for HMXBs in M33 within the PHATTER survey footprint.

\citet{Lazzarini2022} measured the spatially resolved recent (back to $\sim$630 Myr) SFH of M33 with color-magnitude diagram (CMD) fitting. They performed CMD fitting with the software package \verb!MATCH! \citep{Dolphin2002} using the optical-only (F475W and F814W bands with central wavelengths of 4750 and 8353 \AA, respectively) PHATTER catalog as input. The optical-only PHATTER catalog covers an area of $\sim$38 kpc$^{2}$. To derive the spatially resolved SFH, the area of the optical-only PHATTER catalog was divided into 2005 roughly 100 pc by 100 pc (24 $^{\prime \prime}$ on a side) regions for which the SFH was measured independently.

The SFH was measured from log(t $yr^{-1}$) of 6.6 to 10.15 with a step size of 0.1. Their CMD fits focused on main sequence stars, optimizing for recovery of the recent SFH. They excluded older, redder stars from their fits, including the vast majority of red giant branch and red clump stars. The recent SFH is reliable back to $\sim$630 Myr, which more than includes the expected main-sequence lifetimes of stars with masses greater than or equal to 8 M$_{\odot}$, which are the focus of our analysis in this paper.

The SFR measurements from the \citet{Lazzarini2022} include errors that were measured using hybrid Monte Carlo (MC) fitting and represent the range of all SFHs that could recreate the observed CMD in that spatial region of M33. As described in Section 3.2 in \citet{Lazzarini2021}, this fitting can result in measurements of 0 SFR in certain time bins with upper errors, creating asymmetrical errors on SFR measurements, resulting in asymmetrical errors on our age distribution. 

We determine the age distribution for our HMXB candidate sample using the spatially resolved recent SFH within the last 80 Myr in the region of the SFH maps containing each of our HMXB candidates. For the region containing each HMXB candidate, we calculate the stellar mass formed in each age bin back to 80 Myr by multiplying that bin's measured SFR by its width in years. We then divide this stellar mass formed in each time bin by the total stellar mass formed within the last 80 Myr its local region of the SFH maps to get a probability distribution for its age. We sum these probability distributions for all HMXB candidates to get an age distribution for the population, which we show for various sub-samples of our HMXB candidate sample in the plots in the left column of Figure \ref{fig:age_prod_rate}. 

We perform the same analysis for an equal number of regions from the \citet{Lazzarini2022} SFH maps. We randomly select regions from the maps with measured star formation within the last 80 Myr. This allows us to compare the age distribution for regions containing HMXB candidates against the overall age distribution for regions across the galaxy. We perform this random selection of regions 10,000 times and plot the median age distribution as a green line in Figure \ref{fig:age_prod_rate}.  We list the identification number of the region in the \citet{Lazzarini2022} SFH maps that contains each HMXB candidate in Table \ref{table:beast_table} so that the SFH can be retrieved for any individual HMXB. We also include the median age of the stars in the region of the SFH maps containing each HMXB candidate that formed in the last 80 Myr in Table \ref{table:beast_table}. The median age is the age at which 50\% of the cumulative stellar mass in that region was formed, with errors showing the ages at which 16\% and 84\% of the cumulative stellar mass was formed.

\subsection{Determining Best HMXB Candidate Sample}\label{sec:flags}
One major goal in our analysis is to identify a clean sample of HMXB candidates in M33 for which both HST and Chandra data are available. We want to remove any non-HMXB sources including background galaxies, supernova remnants, and foreground Milky Way stars. We assess the quality of our HMXB candidates using each source's X-ray and ultraviolet/optical/infrared properties, which we discuss in more detail in the following subsections.

We use a system of flags to assess the quality of each HMXB candidate. In this system, no single flag completely removes a source from our HMXB candidate sample, but we note that it may be more likely to be a non-HMXB source. We present the values used to evaluate each source in Table \ref{table:summary_table_values} and list each source and the flags that it raised in Table \ref{table:summary_table_flags}. 

\subsubsection{Flag: Soft \textit{Chandra} Hardness Ratios}
HMXBs are known to have hard X-ray spectra, so we can use a source's hardness ratio from the \citet{Tullmann2011} \textit{Chandra} catalog to flag sources with soft hardness ratios, suggesting that they may not be HMXBs. We use the two hardness ratios defined in \citet{Plucinsky2008}: $ HR_{1} = \frac{M-S}{S+M+H}$ and $ HR_{2} = \frac{H-M}{S+M+H}$, where $S=0.35-1.1$ keV, $M=1.1-2.6$ keV, $H=2.6-8.0$ keV. To meet the criteria for a soft \textit{Chandra} hardness ratios flag, a source must have $HR_{1}<-0.4$ and $HR_{2}<0.1$. These cutoff values were determined using the hardness ratio diagrams in \citet{Plucinsky2008} and are intended to flag potential foreground stars and supernova remnants. 

We find that two of the HMXB candidates in our sample raise the soft hardness ratio flag (ChASeM33 IDs 363, 542) suggesting that they are more likely to be foreground stars or supernova remnants.

\subsubsection{Flag: Foreground IR Colors}
Infrared colors can be used to flag potential foreground stars in our sample of HMXB companion star candidates. Foreground stars create a nearly vertical sequence on the F160W versus F110W-F160W CMD with F110W-F160W colors between 0.4 and 0.8 \citep[see Figure 19 in][]{WilliamsPHAT}.

We flag HMXB companion star candidates in our sample that have IR colors in this range, which could suggest that the optical counterparts may be foreground stars, rather than HMXB companion stars in the disk of M33. We emphasize that while a source may have an IR color in this range, that does not definitively indicate that it is a Milky Way foreground star, which is why we use this series of checks to evaluate the quality of our HMXB companion star candidates. We find that 15 stars in our HMXB companion star candidate sample raise the IR color flag. We list the F110W-F160W color for each HMXB companion star candidate in Table \ref{table:summary_table_values} and denote whether or not each HMXB companion star candidate raised the ``IR color flag'' in Table \ref{table:summary_table_flags}.

\subsubsection{Flag: Flat SED Shape}
Background AGN are known to have fairly flat spectra in the range of optical wavelengths that are covered by our photometric measurements from PHATTER \citep{Brown2019} while we expect a star to have a more curved, black-body type spectrum. When we run the BEAST, it creates SED plots for each source showing the flux at the central wavelength of each filter included in the input photometry. Two co-authors visually inspected these SEDs and flagged sources for which the SED shape appears flat. We note that this flattening of the spectrum could be due to large amounts of dust extinction on a bright blue star and thus re-emphasize the fact that we do not rule out a HMXB companion star candidate if it raises just one flag. We find that the sources for which the SED shape appears flat tend to have higher best-fit A$_{V}$ values, between A$_{V}$=3.6 and 4.6.

We find nine HMXB companion star candidates that have flat SED shapes, which have a value of 1 listed in the ``FLAG: Flat SED Shape'' column in Table \ref{table:summary_table_flags}.

\subsubsection{Flag: No Recent SF}
Due to their young ages, we expect HMXBs to be in regions of the galaxy that have had active star formation within the time scale of the lifetimes of massive stars, within the last $\sim$ 80 Myr. For each source, we identify the region of the M33 spatially resolved recent SFH maps \citep{Lazzarini2022} that include the X-ray source position and flag sources where the surrounding 100 pc by 100 pc region has a best-fit SFH consistent with zero star formation in the last 80 Myr. This could indicate that our source is not actually a HMXB, which we would expect to be spatially correlated with recent SF, but could instead be a background galaxy or foreground star, neither of which should have positions correlated with SF regions in M33. 

One of the HMXB candidates in our sample is located in regions of M33 with 0 measured star formation within the last 80 Myr (ChASeM33 ID 205). We list the mean SFR over the last 80 Myr for the region containing each source in Table \ref{table:summary_table_values}.

\subsubsection{Flag: Mis-matched N$_{H}$ and A$_{V}$}
For HMXBs in the disk of M33, we expect the absorption measured in the X-ray spectrum, N$_{H}$, to correlate with the absorption we measure for the companion star at optical wavelengths, A$_{V}$, because the X-ray source and optical source are located within the same binary stellar system. We can use this assumption to flag sources that may not be HMXBs in the disk of M33: i.e. sources where the N$_{H}$ measured for the X-ray source does not correlate with the measured A$_{V}$ for the optical counterpart. Not all X-ray sources we identify as HMXB candidates have measured N$_{H}$ values, thus we only apply this flag to sources for which this measurement is available.

We adopt N$_{H}$ values for sources from \citet{Tullmann2011}. They performed spectral fits for sources with a S/N of at least 2.0 in each bin and a minimum of 8 spectral bins. The best-fit model for each source is listed in Table 9 of \citet{Tullmann2011}. The N$_{H}$ value listed for each source in \citet{Tullmann2011} is the intrinsic N$_{H}$ in M33, with a fixed Galactic N$_{H}$ of $0.06 \times  10^{22}$ cm$^{-2}$. We list the N$_{H}$ value for each of our HMXB candidates for which a fit was performed in Table \ref{table:summary_table_values}.

We compare these N$_{H}$ values with measurements of A$_{V}$ from BEAST SED-fits of the companion star candidates and local measurements of A$_{V}$ that were derived by \citet{Lazzarini2022} while measuring the spatially resolved recent SFH. See Section \ref{sec:SED-fitting} for a detailed description of the SED fitting. \citet{Lazzarini2022} measured the foreground extinction, A$_{V}$, and differential extinction, dA$_{V}$ by fitting optical color-magnitude diagrams over a grid of A$_{V}$ and dA$_{V}$ values, using maximum likelihood estimation.

To compare the X-ray and optical extinction, we use published relationships between N$_{H}$ and A$_{V}$ \citep{predehl1995,Guver2009} to calculate the A$_{V}$ that would be expected for a source given the N$_{H}$ measured in its X-ray spectrum. We then compare this expected A$_{V}$ against the A$_{V}$ we measure with SED-fitting and from the SFH maps. 

We flag sources for which the A$_{V}$ expected for the source based on its measured N$_{H}$ differs significantly from the measured A$_{V}$ from SED-fitting and from the SFH maps by more than 3 times the 90\% confidence errors reported on the N$_{H}$ measurements. For sources without reported errors on the measured N$_{H}$, we assumed 10\% errors. We perform this calculation with both the \citet{predehl1995} and \citet{Guver2009} correlations and find that the outliers are the same for both sets of calculations.

We only find one source that is a clear outlier for which the A$_{V}$ expected based on the X-ray source's N$_{H}$ exceeds the A$_{V}$ measured with SED-fitting of the companion star candidate and the A$_{V}$ measured for the area surrounding the source from the SFH maps. This is ChASeM33 ID 207 and its associated optical counterpart candidate. One possible explanation for this mismatch may be that the X-ray source we are observing is not an accreting black hole or neutron star in the disk of M33, but is instead a background AGN viewed through M33's disk. When we fit for the A$_{V}$ with SED-fitting, we assume sources are stars in M33, which could account for the mis-match. We expect that the N$_{H}$ measured for a background galaxy would not match the A$_{V}$ from the SFH maps near the position of the background galaxy on the sky.

We do not find any sources for which the A$_{V}$ expected based on the X-ray source's measured N$_{H}$ is significantly less than the A$_{V}$ we measure for the source with SED-fitting and the A$_{V}$ in the local area of M33 from the SFH maps. A mis-match of this type could suggest that the X-ray source we are looking at is a foreground Milky Way star rather than an HMXB in the disk of M33. As we described for the previous set of mis-matches, the fact that our SED-fits report a higher A$_{V}$ could be due to the fact that we fix the distance of the companion star candidate to M33 when we perform the fitting. Foreground stars should be very bright and red, which the BEAST could interpret as being a massive star in M33 with significant dust extinction.

\subsubsection{Flag: Spectral Type not consistent with massive star}
In Section \ref{sec:SED-fitting} we describe how we inferred the most likely spectral type for our HMXB companion star candidates based on their best-fit effective temperature, luminosity, and radius. In Figure \ref{fig:HR_diagram} we plot the values we fit for each star along with the expected ranges in effective temperature and luminosity for massive main sequence, giant, and supergiant stars, which we also list in Table \ref{table:sp_type_ranges}. We flag HMXB  companion star candidates for which the best-fit effective temperature, luminosity, and radius do not fit into the expected range for O or B type main sequence, giant, or supergiant stars. We find five HMXB companion star candidates that raise this flag. This could indicate that the point source for which we are performing the SED fitting is not a star in M33, which could mean that it is a foreground Milky Way star or a background galaxy viewed through the disk of M33. 

\subsection{Cross-Match with Previous Catalogs}\label{sec:cross-match}
We cross-match our sample with previously published X-ray catalogs of M33 \citep{Grimm2005,Williams2015,Yang2022} and previous catalogs of HMXB candidates in M33 \citep{Garofali2018}.

We identify counterparts to our HMXB candidates in the \citet{Yang2022} NuSTAR survey of M33. NuSTAR detected sources in that paper are listed with their counterpart IDs in the ChASeM33 survey \citep{Tullmann2011}, which we use to identify counterparts between the two catalogs. To identify counterparts in the \citet{Williams2015} XMM-Newton catalog and the \citet{Grimm2005} Chandra catalog, we positionally cross-match sources within 5$^{\prime \prime}$. We list the ID number of the counterparts in the \cite{Yang2022}, \citet{Williams2015}, and \citet{Grimm2005} catalogs in Table \ref{table:beast_table}.

\subsubsection{Comparison with Garofali et al. (2018) HMXB Sample}
We performed a detailed cross-check against the HMXB candidate sample presented in \citet{Garofali2018}, which also used the ChASeM33 catalog to identify HMXB candidates using overlapping optical data from archival HST fields. Because of the non-uniform coverage of the HST fields used in their analysis, not all HMXB candidates identified by \citet{Garofali2018} fall within the PHATTER survey footprint, and thus are not included in this analysis. 

\citet{Garofali2018} presented 55 HMXB candidates, 40 of which lie in the footprint of the PHATTER survey footprint and 15 of which lie outside the footprint of the PHATTER survey. We identify 15 HMXB candidates in this study that fall within the footprint of the HST observations analyzed by \citet{Garofali2018} but that are not included in their catalog. Many of the HST observations used by \citet{Garofali2018} were shallower than the PHATTER data used here, which may account for some of this discrepancy. Additionally, many of the HST fields used in their analysis did not cover the full photometric range of the data we use here, and many did not have coverage at UV wavelengths, which we used to identify HMXB companion star candidates. 

There are 8 HMXB candidates identified by \citet{Garofali2018} that fall within the PHATTER footprint but that are not included in our HMXB candidate sample. Two of these X-ray sources (ChASeM33 sources 210, 337) do not fit our selection criteria and did not have the same photometric coverage in the data analyzed by \citet{Garofali2018}. There are 4 HMXB candidates (328, 357, 362, 427) that were included by \citet{Garofali2018} for completeness, but that we do not include in our sample because they were classified as supernova remnants by \citet{Long2010}. Source 416 is identified as an HMXB candidate by \citet{Garofali2018} but we do not include it in our sample because there is a resolved background galaxy at the location of the X-ray source in the PHATTER imaging. We do not include source 535 in our sample, although \citet{Garofali2018} included it, as it was later classified as a colliding-wind binary by \citet{Garofali2019}.

The remaining 32 HMXB candidates identified by \citet{Garofali2018} that fall within the footprint of the PHATTER survey were independently identified as HMXB candidates in this analysis. In Table \ref{table:beast_table} we include a column that indicates which sources are a match between the two catalogs and list the ChASeM33 name used to identify each source in the \citet{Garofali2018} catalog for matches.

\section{Results \& Discussion}\label{sec:results}
In this section we discuss the SED-inferred physical properties of the HMXB companion star candidates in our sample, the age distribution and HMXB production rate we measure, compare the characteristics of the M33 HMXB population with those measured for the population of HMXBs in M31, which was studied with an analogous approach in \citet{Lazzarini2021}, and discuss the hard X-ray properties of HMXB candidates in our sample that were also detected in a recent NuSTAR survey of M33 \citep{Yang2022}.

\subsection{SED-inferred properties of HMXB Companion Star Candidates in M33}
As described in Section \ref{sec:SED-fitting}, we classify HMXB companion star candidates as O- or B-type main sequence, giant, or supergiant stars based on their best-fit effective temperature, luminosity, and radius. There are several known classes of HMXBs, based on the nature of the companion star and the accretion mechanism. Supergiant HMXBs (sgHMXBs) are systems where the black hole or neutron star accretes material from a supergiant companion star, either via stellar winds or Roche lobe overflow. In Be X-ray binaries (Be-XRBs), the companion star is an Oe/Be star, which is rapidly rotating and forms an equatorial decretion disk that exhibits characteristic Hydrogen emission lines in the star's spectrum. Accretion onto the compact object occurs when the compact object, typically a neutron star, passes through or very close to this decretion disk. The Oe/Be companion stars in Be-XRBs have been observed to have spectral types ranging from late O-type stars (O9) to early B-type stars \citep[see][for a review]{Reig2011}. 

Classifying the companion star type in HMXB systems is only possible for nearby galaxies and the Milky Way, where optical spectroscopy is readily available. In the Milky Way sgHMXBs make up roughly 30\% of the observed HMXB population, Be-XRBs make up about 50\%, and the remaining 20\% is a combination of giant, main sequence, and unknown companion star types. In contrast, in the Magellanic clouds, out of over 120 HMXB systems, all but two are known to contain Be companion stars \citep{Liu2006,Maravelias2014,Walter2015,Haberl2016,yang2017multi}.

We can attempt to compare the breakdown of likely spectral types of the HMXB companion stars in our observed sample with the populations in the Milky Way and Magellanic Clouds. However, we note that this is challenging because 23 of our HMXB candidates have more than one candidate companion star.

In systems with only one potential HMXB companion star candidate, we find that 31 (76\%) are likely B-type main sequence stars and 2 (5\%) are likely O-type main sequence stars. Of the remaining systems, 1 is a likely B-type giant, 2 are likely O-type giants, and 2 are likely B-type supergiants.

In systems with more than one candidate companion star, we find that 14 (61\%) only have B-type main sequence stars as potential companions (i.e. all candidate companion stars are likely B-type main sequence stars). The remaining systems with multiple candidate companion stars have a mix of O and B-type main sequence, giant, and supergiant stars. 

Because there are multiple candidate companion stars for many of our HMXB candidates, we cannot state a firm fraction of the sample that host different spectral type companion stars. However, because most of the systems contain a likely O- or, more frequently, B-type main sequence companion star, our findings suggest that most of the HMXBs in M33 are Be-XRBs. Future optical spectroscopy is required to confirm the true spectral type of the companion stars in these systems, but these initial findings suggest that the fraction of Be-XRBs for M33 lies somewhere between the fraction in the Milky Way and the fraction in the Magellanic Clouds. Metallicity has been suggested to affect the fraction of giant/supergiant HMXBs to Be-XRBs in a population of HMXBs, due to lower rates of mass loss via line driven winds at lower metallicities \citep{Linden2010}. 

We note that our SED fits use single stellar models and these models do not include a component to model the potential decretion disk around Be stars. Be stars exhibit infrared excesses compared to normal O/B type stars. This effect is orientation dependent, and is most significant when the system is viewed pole-on. This excess has a small ($<$0.5 mag) effect at optical wavelengths but can dominate at infrared wavelengths \citep[e.g.][]{Rivinius2013}. We include the two optical and two ultraviolet bands from the PHATTER photometric catalog in our SED-fits, so we do not expect infrared contamination from the decretion disk to affect our measurements. There is a chance that we get contamination in the ultraviolet bands from the compact object's accretion disk, which may bias our SED fits towards higher masses. This is a challenging system to disentangle with purely photometric data, and we plan to tackle this with forthcoming spectroscopic measurements (Lazzarini et al., in preparation).

\subsection{Age Distribution of HMXBs in M33}\label{sec:age_distribution}
As described in Section \ref{sec:age_determination}, for each HMXB candidate, we used the SFH within the last 80 Myr in the 100 pc by 100 pc region surrounding the source to calculate a probability distribution for its age. We performed this calculation for all HMXB candidates in our sample and then sum their probability distributions to create an age distribution for the population. The sub-samples represented in each row of Figure \ref{fig:age_prod_rate} reflect the HMXB candidates that raise any number of flags, at most one flag, and zero flags in the top, middle, and bottom rows, respectively. We include the known nuclear HMXB (X-8) in our best HMXB candidate sample although we cannot identify its optical counterpart due to the high density in the nuclear cluster of M33, which brings the total number of HMXBs in our full sample to 65.

We present this age distribution, for different sub-samples, with the black histogram in the plots in the left column of Figure \ref{fig:age_prod_rate}. The left plot in the top row of Figure \ref{fig:age_prod_rate} shows the age distribution for our full sample of HMXB candidates (65 sources), the middle row shows the sub-sample that raised one or fewer flags (63 sources), and the bottom row shows our highest quality sub-sample that raised 0 flags (53 sources), using the evaluation criteria described in Section \ref{sec:flags}. We include error bars on the age distribution for each sub-sample. We calculate the error bars by generating a PDF for each source from a random sampling within the errors on its SFR measurements. We performed this random selection for all sources in the sub-sample 10,000 times and present the 16th and 84th percentile of the age distribution derived from this sampling as the lower and upper errors on the age distribution, respectively. 

On each age distribution plot, we also include a green line which shows the results when we perform the same analysis on an equal number of randomly selected regions from the SFH maps that have a measured SFR$>$0 M$_{\odot}$ yr$^{-1}$ over the last 80 Myr. We run this random selection 10,000 times and the green line represents the median age distribution, which is a proxy for the overall SFH in M33.

As shown in the bottom row, left panel of Figure \ref{fig:age_prod_rate}, we find more HMXB candidates associated with star formation between 0 and 10 Myr ago and between 40 and 50 Myr ago than would be expected based on the green random line, which represents sampling from the SFH maps agnostic of the positions of HMXB candidates. This was also seen in a previous study of the HMXB population of M33 by \citet{Garofali2018} where the age distribution of HMXBs in M33 was measured using SFHs derived with CMD-fitting of archival HST observations. \citet{Garofali2018} suggested that these peaks represent two different HMXB formation channels: a prompt channel (0-10 Myr) and a more delayed formation channel (40-60 Myr). 

Different types of HXMBs have been shown to form via different channels, with associated timescales \citep[e.g.;][]{Linden2009,Linden2010}. The 40$-$50 Myr peak in the HMXB production rate that we observe is the expected formation timescale for Be-XRBs. The 0$-$10 Myr peak in the HMXB production rate we observe is characteristic of more massive binaries, where the primary star forms a compact object and starts accreting within a few Myr. Previous studies in other galaxies including the SMC \citep{Antoniou2009}, NGC 300, and NGC 2403 \citep{Williams2013}, found that the preferred age of HMXBs in those galaxies is $\sim$40$-$55 Myr. In M31 the preferred ages for HMXBs is $\sim$10$-$50 Myr \citep{Williams2018,Lazzarini2018,Lazzarini2021}, a broader range that overlaps with the findings in the SMC, NGC 300, and NGC 2403. In the LMC, by contrast HMXBs are associated with younger star formation episodes. The SFH is peaked at $\sim$6.3 Myr in the regions surrounding sgHMXBs and black hole HMXBs, $\sim$12.6 Myr in the regions surrounding Be-XRBs and X-ray pulsars, and $\sim$25 Myr in the region surrounding a candidate white dwarf-Be X-ray binary \citep{Antoniou&Zezas2016}. Our findings suggest that we are observing an HMXB population in M33 that is dominated by two different formation channels with timescales characteristic of Be-XRBs (40$-$50 Myr) and more massive HMXBs (0$-$10 Myr).

\subsubsection{Masses and Ages}\label{sec:masses_ages}
To better understand formation channels for our observed HMXB population, we explored the relationship between the mass of the HMXB companion star candidates that we infer from SED-fitting as described in Section \ref{sec:SED-fitting} and the ages of the systems measured using the SFH maps from \citet{Lazzarini2022}.

In Figure \ref{fig:age_mass} we plot the HMXB companion star candidate mass against two measures of likely age for the system. In both panels points are color-coded according to the likely spectral type of the companion star. Diamonds indicate HMXBs for which there is only one companion star candidate, while squares represent HMXBs with multiple companion star candidates. The opacity of each point scales with its quality as defined in Section \ref{sec:flags}. The gray line in the background shows the maximum stellar mass expected at each age in single stellar evolution, using Padova isochrones \citep{Marigo2008}. We also include outlines around points for HMXB candidates associated with a compact object with a classification based on its NuSTAR colors and luminosity, as described in \citet{Yang2022} and listed in Table \ref{table:beast_table}. Circles indicate sources that are likely black holes (in any accretion state) and stars indicate sources that are likely neutron stars (either pulsars or lower magnetic field neutron stars).

In the right panel we plot the median age of stars that formed in the last 80 Myr in the 100 pc by 100 pc region of the SFH maps that contains each HMXB candidate. The median age is defined as the age at which 50\% of the total stellar mass in the region has formed, and we include error bars that extend to the ages at which 16\% and 84\% of the stellar mass was formed, respectively. The median age is not generally a good representation of the likely age of the HMXB candidate because of contamination from older stars that are not necessarily associated with the star formation event that formed the HMXB candidate. As Figure \ref{fig:age_prod_rate} shows, some regions may have both an early and a later peak in their SFH, which would cause the median age to be somewhere in between these two peaks, when the SFR was actually lowest and the systems are less likely to have formed.

We looked at the age of the most recent significant star formation to try and remove contamination from older star formation episodes in the surrounding region that are not associated with the star formation episode that produced the HMXB. We defined significant star formation as a SFR that would produce 1000 M$_{\odot}$ during the 10 Myr time bin, which assuming a \citet{Kroupa2001} initial mass function, should produce $\sim$10 stars with M$>$8 $M_{\odot}$. This threshold SFR is greater than the noise in the SFR measurements in the SFH maps, which is $\sim$1.0$\times10^{-5}$ M$_{\odot}$ yr$^{-1}$ after re-binning the spatially resolved recent SFH maps from \citet{Lazzarini2022} into 10 Myr time bins.

As is shown in Figure \ref{fig:age_mass}, the age of the most recent significant star formation correlates with the mass of the companion star derived via SED-fitting, although these two measurements are completely independent. We see that systems containing our most massive HMXB companion star candidates are found in regions of M33 with very recent (between 0 and 20 Myr ago) significant SF. We find that for HMXB companion star candidates reside in regions with less recent episodes of significant star formation, the masses of the companion stars are lower.

There are two outliers from this relationship with ages of most recent significant star formation older than expected given the SED-derived mass of the HMXB companion star candidate. These systems could indicate an extension of the stellar lifetime due to binary interactions \citep[e.g.,][]{Zapartas2017} or a system with an abnormally high velocity, but additional characterization of these HMXB systems is needed. We have an ongoing program to obtain optical spectroscopic observations of the M33 HMXB companion star candidates presented in this paper, which is needed to confirm the spectral types, and thus masses, of the HMXB companion star candidates (M. Lazzarini, et al., in preparation).

\subsection{HMXB Production Rate}\label{sec:hmxb_prodrates}
The production rate of HMXBs in M33, or the number of HMXBs produced per unit stellar mass or unit SFR, can be compared with measurements from other galaxies to explore the relationship between HMXB production and other galaxy-wide properties such as metallicity, stellar mass, and star formation rate. We present two measures of HMXB production rate in the following subsections. First we present the HMXB production rate measured in the number of HMXBs produced per unit SFR (HMXBs/(M$_{\odot}$ yr$^{-1}$)). We also measure a time-resolved HMXB production rate, or the number of HMXBs produced per stellar mass, over the last 80 Myr. We provide more details on each calculation in the following subsections and discuss the HMXB production rate in M33 in comparison with other galaxies for which this measurement has been done.

Our HMXB production rate measurements are a reflection of the number of HMXBs that were active at the time of the ChASeM33 survey Chandra observations. This number is impacted by the duty cycle for HMXBs, or the fraction of time that they spend in the active HMXB phase. Empirical studies of mostly Galactic HMXBs have revealed duty cycles ranging from around 10\% (for transient Be-XRBs) \citep{Sidoli2018} to greater than 10\% (for SG HMXBs), which means that our HMXB production rates are a lower limit on the true HMXB production rate because a significant fraction of HMXBs in M33 were likely not active at the time of observation. 

We calculate the HMXB production rate per unit SFR over 50 and 80 Myr timescales. 50 Myr is the expected maximum age for massive stars from single stellar evolution, however this lifetime may be extended for stars in binary systems due to rejuvenation from mass transfer via binary interactions \citep[e.g.,][]{Zapartas2017}.

\subsubsection{Calculation of HMXBs per Unit SFR}
We can use calculations of the HMXB production rate per unit SFR to compare with measurements in other galaxies where time-resolved SFH measurements are not available. We calculate the number of HMXBs formed per unit SFR (HMXBs/M$_{\odot}$ yr$^{-1}$) over the last 50 and 80 Myr, to account for the lifetimes of single massive stars and potential extension of that lifetime due to binary stellar interactions.

To calculate the number of HMXBs formed per unit SFR, we integrate the age distribution of HMXBs shown in the left column in Figure \ref{fig:age_prod_rate} over 50 and 80 Myr. We then divide this number of HMXBs by the mean SFR over the full PHATTER footprint over the same timescale. The whole PHATTER footprint is covered by the ChASeM33 survey, which means that we are sensitive to detecting HMXBs over its full area. We obtain a lower and upper limit on this measurement from the error bars on the SFR measurements. We present the HMXB production rates per unit SFR for various sub-samples in Table \ref{table:HMXB_prod_rates}.

We aim to compare this HMXB production rate with rates measured in other galaxies. However, this direct comparison is difficult due to differences in the measurement of SFR in these galaxies and the way that HMXB samples were generated including the sensitivity of X-ray observations. Due to their proximity, M33 and M31 both has spatially resolved SFH maps. The LMC and SMC also have spatially resolved SFH maps \citep{Zaritsky&Harris2004,Zaritsky2009} that have been used to associate their HMXBs with star formation episodes, and thus ages and production rates \citep{Antoniou2010,Antoniou&Zezas2016,Antoniou2019}. 

\citet{Politakis2020} performed a direct comparison of the HMXB production rates of NGC 55, the LMC and SMC, and the Milky Way using the HMXB candidate samples presented in \citet[SMC, LMC;][]{Antoniou&Zezas2016}, \citet[Milky Way;][]{Bodaghee2012} and \citet[NGC 55;][]{Harris&Zaritsky2009}. They used the number of known HMXB candidates in each galaxy and the mean SFR of each galaxy to calculate their global HMXB production rates. These HMXB production rates show a general trend in increasing HMXB production rate with decreasing metallicity. The lowest metallicity galaxies in their sample, NGC 55 and the SMC, have HMXB production rates of 181$-$240 and 240$-$880 HMXBs/(M$_{\odot}$ yr$^{-1}$). They measure an HMXB production rate of 96$-$256 and 52$-$86 HMXBs/(M$_{\odot}$ yr$^{-1}$) in the LMC and Milky Way, respectively. For our full HMXB candidate sample, we measure an HMXB production rate of 181$-$240 HMXBs/(M$_{\odot}$ yr$^{-1}$) over the last 80 Myr, which places the HMXB production rate of M33 higher than that of the Milky Way, similar to the HMXB production rate of the LMC, and lower than the measured rate in the SMC and NGC 55.

\subsubsection{Calculation of Time-Resolved HMXB Production Rate}
\begin{figure*}
\centering
\includegraphics[width=0.8\textwidth]{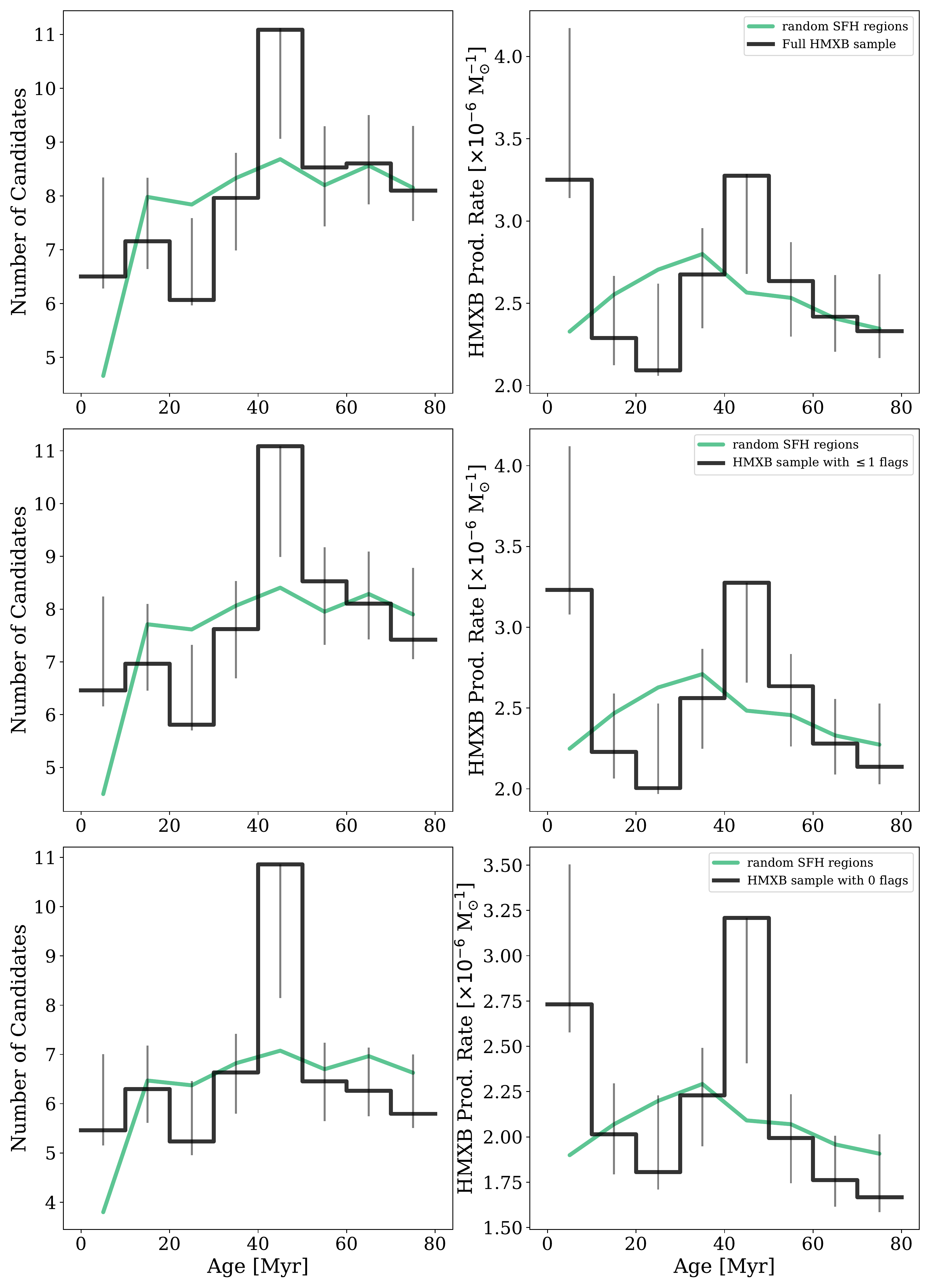}
\caption{\small{Age distributions and time-resolved production rates for our HMXB sample. Each row represents a different sub-sample based on the evaluation criteria described in Section \ref{sec:flags}, the top row is our full sample, the middle row represents HMXB candidates that raise one or fewer of our quality flags, and the bottom row shows our highest quality HMXB candidate sample. The black histograms in the left column represent the age distribution for each sub-sample and the green line represents a random sample. For more detail, see Section \ref{sec:age_distribution}. In the right column, the histograms represent the time-resolved HMXB production rate for each sample and the green line represents a random sample for comparison. See Section \ref{sec:hmxb_prodrates} for more details.}}
\label{fig:age_prod_rate}
\end{figure*}

\begin{figure*}
\centering
\includegraphics[width=0.95\textwidth]{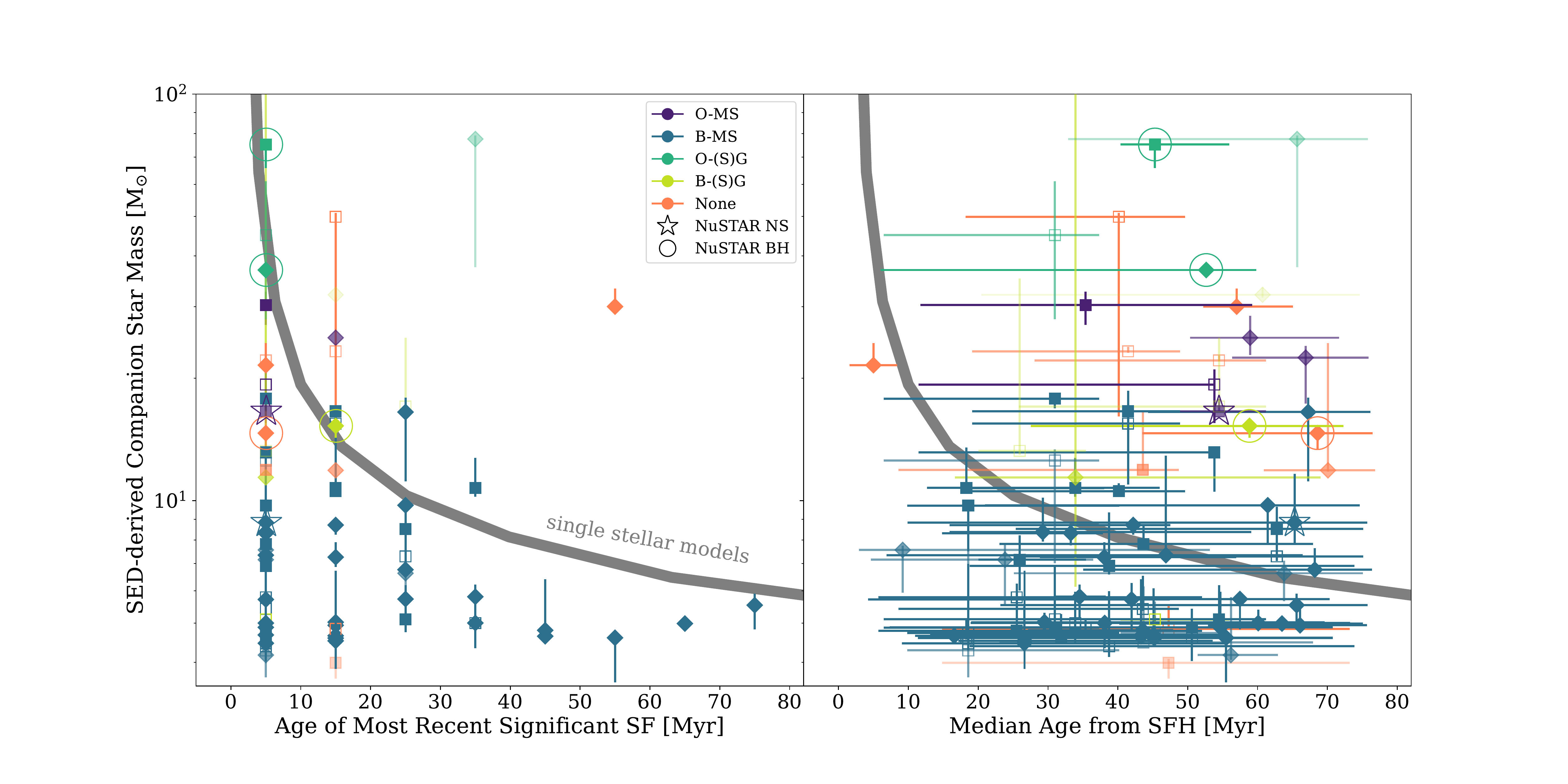}
\caption{\small{In this figure we compare the mass of the HMXB companion star candidates derived via SED-fitting and the likely ages of the systems derived from the spatially resolved recent SFH map of M33 \citep{Lazzarini2022}. Additional details about this figure can be found in Section \ref{sec:masses_ages}. \textbf{Left}: Each point represents one HMXB companion star candidate, listed in Table \ref{table:beast_table}. On the y-axis we plot the mass of the HMXB companion star candidate derived via SED-fitting. On the x-axis we plot the age bin (i.e. we plot a point at 5 Myr for an age bin of 0$-$10 Myr) of the most recent significant star formation in the SFH map region including the HMXB candidate. We define significant star formation in more detail in Section \ref{sec:masses_ages}. Points are color-coded according to the likely spectral type of the companion star. Diamonds indicate HMXBs for which there is only one companion star candidate, while squares represent HMXBs with multiple companion star candidates. The opacity of each point scales with its quality as defined in Section \ref{sec:flags}. The gray line in the background shows the maximum stellar mass expected at each age in single stellar evolution. We outline HMXB candidates with compact object classifications from NuSTAR observations, described in Section \ref{sec:nustar}. \textbf{Right:} We plot the HMXB companion star candidate masses on the y-axis and the median age of stars formed in the last 80 Myr in the SFH map region including the HMXB companion star candidate. Point colors and shapes follow the same conventions as the left panel.}}
\label{fig:age_mass}
\end{figure*}

We can also calculate a time-resolved measure of the HMXB production rate in M33 using the measured HMXB age distribution and the SFH of M33 in the PHATTER footprint. The age distribution indicates the number of HMXB candidates that likely formed in each 10 Myr time bin. To derive the number of HMXBs produced per stellar mass, we must first calculate the total stellar mass formed in each 10 Myr time bin in the area covered by both the PHATTER and ChASeM33 surveys in M33. We find the total stellar mass formed in each time bin by multiplying the SFR in each time bin by its width, 10 Myr. We can then divide the HMXB age distribution by the stellar mass formed in each time bin to produce a measurement of the number of HMXBs formed per stellar mass. We present this time-resolved HMXB production rate in the right column of Figure \ref{fig:age_prod_rate}, with the full sample of HMXBs on the top row, the sample of HMXBs which raise one or fewer flags in the middle row, and our highest quality sample of HMXBs in the bottom row. The green line in the right column of Figure \ref{fig:age_prod_rate} indicates the same analysis performed with an equal number of randomly selected regions from the \citet{Lazzarini2022} SFH maps with measured SFR over the time range analyzed (mean SFR over the last 80 Myr $>$0).

\subsection{Comparison with M31 HMXB Population}
We can compare the HMXB populations of M33 with the HMXB population of M31 to understand how differences between their histories of star formation are reflected in the HMXB populations we observe today. Previous studies of the HMXB population of M31 used the PHAT dataset and overlapping Chandra-PHAT survey \citep{Williams2018} to identify HMXB candidates, characterize the companion stars with SED-fitting, measure their age distribution and HMXB production rates with spatially resolved recent SFH maps \citep{Lazzarini2018,Williams2018,Lazzarini2021}. The PHAT and PHATTER surveys were designed to be very similar, making the population demographics of their HMXB populations inferred from these surveys directly comparable. Both the \citet{Lazzarini2021} study of M31 and this one use Chandra observations with similar depths.

We find that the HMXB populations in M31 and M33 are remarkably similar. Despite known differences in the area of the surveys and variation in the SFH of each galaxy, we find a roughly equal number of HMXB candidates in each study. We identify 65 HMXB candidates in our full sample in M33 and \citet{Lazzarini2021} identified 57 HMXB candidates in M31 using very similar methodology. While at first glance, this may seem surprising, upon closer inspection this similarity is supported by known correlations between SFR, metallicity and the X-ray luminosity function.

While M33 is known to exhibit a decreasing metallicity gradient with radius, at the radii probed by the PHATTER survey in M33 and the PHAT survey in M31, the gas-phase metallicities of these galaxies are comparable. The gas-phase metallicity of M33 within the PHATTER footprint ranges from $12+log(O/H)\sim8.4-8.7$ while within the M31 PHAT survey footprint, the gas-phase metallicity ranges between 8.6 and 9.0 \citep[][and references therein]{Williams2021}. While the luminosity of HMXBs is known to scale with metallicity, this effect is dominated by high luminosity X-ray sources (L$_{0.5-8.0keV}>10^{38}$ erg s$^{-1}$) and below this limit, the X-ray luminosity function has been shown to be a nearly universal power-law distribution \citep{Lehmer2021}. There is only one identified HMXB in M31 and M33 with a measured L$_{0.5-8.0keV}>10^{38}$ erg s$^{-1}$ --- M33 X-8 --- which suggests that the difference we see in the HMXB populations cannot be explained by metallicity alone.

The total SFR measured in the areas of M31 and M33 covered by PHAT/Chandra-PHAT and PHATTER/ChASeM33, respectively, are remarkably similar, despite covering significantly different de-projected areas. The area covered by overlapping PHAT and Chandra-PHAT observations in M31 is $\sim$328 kpc$^{2}$, which is almost 9 times larger than the $\sim$38 kpc$^{2}$ area covered by the overlapping PHATTER and ChASeM33 surveys in M33. We summarize the mean SFR in the areas of M31 and M33 covered by both HST and Chandra, used in this analysis and \citet{Lazzarini2021} in Table \ref{table:m31_m33_sfr} over the last 50, 80, and 100 Myr. M33 has a slightly higher SFR on all timescales.

\begin{figure}
\centering
\includegraphics[width=0.45\textwidth]{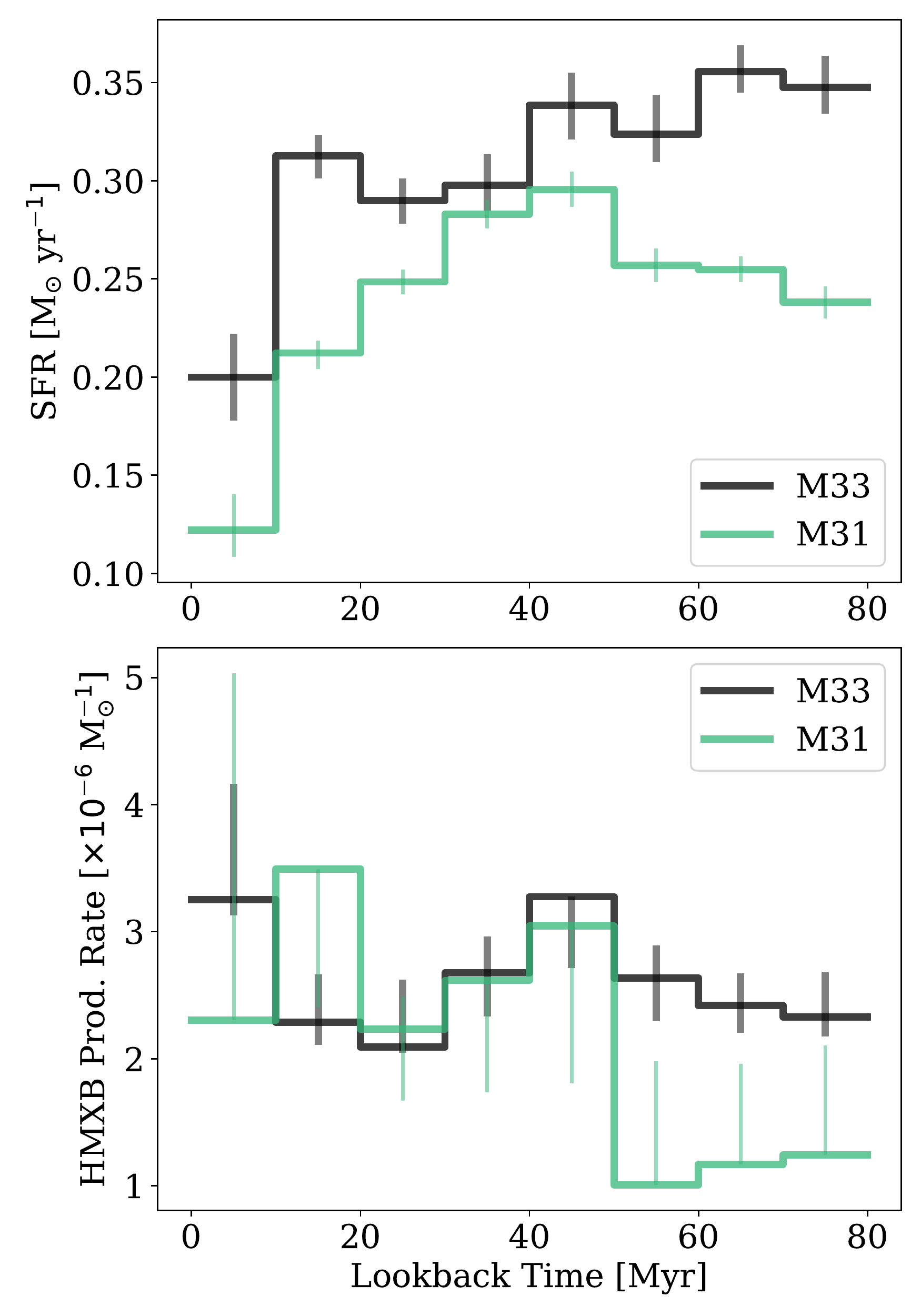}
\caption{Comparison of the SFH in the regions of M31 and M33 covered by overlapping Chandra and HST observations used in this analysis (M33) and in \citet[][; M31]{Lazzarini2021}. \textbf{top:} The SFR over the last 80 Myr in the areas of M31 and M33 covered by the PHAT and PHATTER surveys, respectively, and overlapping Chandra observations. The M31 SFH measurements come from spatially resolved recent SFH maps measured with the optical PHAT photometry \citep{Lewis} and the M33 SFH measurements come from similar maps in M33 produced with the PHATTER photometry \citep{Lazzarini2022}. \textbf{bottom:} The time-resolved HMXB production rate in M31 and M33.}
\label{fig:m31_m33_hmxb_prod_rates}
\end{figure}

It is also interesting to contrast the time resolved HMXB production rate measurements. In Figure \ref{fig:m31_m33_hmxb_prod_rates} we plot the time-resolved HMXB production rates for the full sample of HMXB candidates in M31 and M33 on the same axes. See Section \ref{sec:hmxb_prodrates} for a detailed description of how the time resolved HMXB production rates were calculated in both galaxies. Briefly, to calculate the time resolved HMXB production rate the number of HMXBs expected to have formed in each time bin, from the population age distribution, is divided by the total SFR of the galaxy within the analysis region during the same time span. Because the total stellar mass formed in the region of each galaxy covered by HST and Chandra surveys is used in the calculation of the time-resolved HMXB production rate, and the total stellar mass is derived from the SFH, differences in the SFH of M31 and M33 should not affect the time resolved HMXB production rates. The HMXB production rates in both galaxies agree within errors in most time bins, but M33 has a slightly higher HMXB production rate at ages beyond $\sim$50 Myr.

\subsection{Hard X-ray Properties of M33 HMXB Candidates}\label{sec:nustar}
Comparing the optical and hard X-ray properties of HMXB candidates can be useful for understanding the binary evolution that led to the systems we observe today. Most of the PHATTER survey footprint has also been covered by a hard X-ray survey of M33 with NuSTAR \citep{Yang2022}. In their catalog, \citet{Yang2022} identified the counterpart for each NuSTAR source in the ChASeM33 survey catalog. \citet{Yang2022} detected 28 hard X-ray sources in M33, 7 of which are HMXB candidates identified in this paper. Due to the large point spread function (PSF) of NuSTAR compared to Chandra, there are several NuSTAR-detected sources with multiple associated ChASeM33 sources. Two of our HMXB candidates (ChASeM33 sources 321 and 398) are associated with NuSTAR sources for which there are multiple ChASeM33 counterparts, thus the NuSTAR source may be a blend of multiple ChASeM33 sources.

Hard X-ray observations with NuSTAR can be used to constrain the type and/or accretion state of an accreting compact object. Due to spectral differences between accreting black holes and neutron stars at hard X-ray wavelengths (E$>$10 keV), the hardness ratios and luminosities of accreting compact objects of unknown type can be used to classify them by comparing to Galactic compact objects for which the compact object type/accretion state is known \citep[e.g.,][]{Wik2014,Yukita2016,Vulic2018,Lazzarini2018,Lazzarini2019,Yang2022}. We list the source ID and compact object classification for HMXB candidates in our sample that were detected in the \citet{Yang2022} NuSTAR survey of M33 in Table \ref{table:beast_table}.

While the number of HMXB candidates for which NuSTAR detections and classifications are available is small, we observe some trends when we compare the likely compact object type inferred with NuSTAR and the likely spectral type of the companion star candidate based on its optical/UV photometry. We find that of the 7 matches, the 3 of the 4 that are classified as black holes, soft or intermediate accretion state, have a likely giant or supergiant companion star candidate. The fourth has a best fit SED-derived luminosity, effective, temperature, and radius that is consistent with being an B supergiant within errors (source 299). One of the matched sources is classified as a ULX and the companion star candidate for this source is a likely B-type main sequence star. The remaining two matches are both classified as neutron stars, one as a pulsar and one as Z-type, and both have likely main sequence type companion stars. 

It is interesting that we find supergiant/giant companion star candidates associated with compact objects classified as black holes based on their NuSTAR colors and luminosities but not neutron stars. Supergiant/giant companion stars have been observed in HMXB systems in other galaxies with both black hole and neutron star accretors. Black hole HMXBs (BH-HMXBs) such as Cyg X-1 \citep{Orosz2011}, LMC X-1 \citep{Orosz2009}, and M33 X-7 \citep[][which is included in our sample as Source ID 225]{Orosz2007,Ramachandran2022}, all comprise black holes accreting from supergiant companion stars with masses $\gtrsim$30 M$_{\odot}$, suggesting they are very young systems because 30 M$_{\odot}$ stars have maximum lifetimes $\leq$ 10 Myr. Neutron star HMXBs (NS-HMXBs) with supergiant companion stars are also thought to be very young systems. The companion stars in these systems have been observed with masses $\gtrsim$15 M$_{\odot}$ and companion star ages of $<$12 Myr \citep[][and references therein]{Reig2016}. It is important to emphasize that the classification of the compact object type based on its NuSTAR colors and luminosity is not conclusive because we do not have dynamical mass measurements for the accreting compact objects in these HMXB systems. More detailed spectroscopic observations are needed to 1. confirm the true spectral types and masses of the companion stars and 2. constrain the compact object masses dynamically. We also do not wish to over interpret this result due to the potential impact of small number statistics in our sample with 4 likely black hole supergiant HMXBs.

\section{Conclusions}\label{sec:conclusion}
We present a study of the HMXB population of M33 using a combination of high resolution X-ray observations from Chandra \citep[ChASeM33][]{Tullmann2011} and near-IR/optical/near-UV observations from HST \citep[PHATTER][]{Williams2021}. We performed SED-fitting on the HMXB companion star candidates to infer their likely spectral types. We used the spatially resolved recent SFH of M33 \citep{Lazzarini2022} to measure the HMXB age distribution and production rate for M33. We discuss our findings in context of previous studies of HMXB populations in nearby galaxies and previous work in M33. We list our main findings below:
\begin{itemize}
    \item We identify 65 HMXB candidates in M33, including X-8 in the nuclear star cluster, using multi-band photometry from the PHATTER survey of the inner disk of M33. 
    \item Using SED-fitting, we infer that the majority of our HMXB companion star candidates are likely B-type main sequence stars, suggesting that a majority of HMXBs in M33 are Be-XRBs, although spectroscopic observations are required to confirm these classifications.
    \item We find that the production rate for HMXBs in M33 is peaked between 0 and 10 Myr ago and between 40 and 50 Myr ago. The 0$-$10 Myr formation timescale is associated with BH-HMXBs with massive companion stars and the $\sim$40 Myr formation timescale is associated with Be-XRBs, we detect signals of these two formation channels in the time-resolved HMXB production rate of M33.
    \item We calculated the time-resolved HMXB production rate (the number of HMXBs formed per stellar mass) in M33 and find that it ranges from $\sim3 \times 10^{-6}$ M$_{\odot}^{-1}$ to $\sim1 \times 10^{-6}$ M$_{\odot}^{-1}$ over the last 80 Myr.
    \item We also calculate the number of HMXBs formed per unit SFR over the last 50 and 80 Myr. For our full HMXB candidate sample, we measure an HMXB production rate of 120$-$153 HMXBs/(M$_{\odot}$ yr$^{-1}$) over the last 50 Myr and 181$-$240 HMXBs/(M$_{\odot}$ yr$^{-1}$) over the last 80 Myr.
    \item We find that the HMXB production rates (HMXBs M$_{\odot}^{-1}$) of M33 and M31 agree within errors over the last 50 Myr and only differ slightly between 50 and 80 Myr ago.
    \item We find that for the 7 HMXB candidates in our sample with compact objects classified as black holes or neutron stars based on NuSTAR observations \citep{Yang2022}, systems with NuSTAR-classified black hole accretors are likely to host giant/supergiant companion stars while systems with NuSTAR-classified neutron star accretors are likely to host main sequence companion stars.
\end{itemize}

\noindent \textbf{Acknowledgements: }Support for this work was provided by NASA through grant \#GO-14610 from the Space Telescope Science Institute, which is operated by AURA, Inc., under NASA contract NAS 5-26555. Margaret Lazzarini was supported by an NSF Astronomy and Astrophysics Postdoctoral Fellowship under award AST-2102721 during this work. The Flatiron Institute is supported by the Simons Foundation.

\begin{longrotatetable}
\setlength{\tabcolsep}{0.7pt}
\begin{deluxetable*}{cccccccccccccccccccccc}
\tabletypesize{\tiny}
\tablecaption{X-ray properties and SED fit parameters for HMXB candidates \label{table:beast_table}}
\tablehead{
\colhead{T11} &
\colhead{T11} &
\colhead{T11} &
\colhead{T11} &
\colhead{PHATTER} &
\colhead{CP} &
\colhead{0.35$-$8 keV} &
\colhead{log(T$_{eff}$)} &
\colhead{log(L)} &
\colhead{R} &
\colhead{A$_{V}$} &
\colhead{M} &
\colhead{SED} &
\colhead{SFH Med.} &
\colhead{L22 SFH} &
\colhead{Best Sp.} & 
\colhead{T11} &
\colhead{G18} &
\colhead{Y22} & 
\colhead{Y22} & 
\colhead{W15} & 
\colhead{G05} 
\\
\colhead{ID} &
\colhead{R.A.} &
\colhead{Dec.} &
\colhead{ePos} &
\colhead{Name} &
\colhead{Sep.} &
\colhead{Lum.} &
\colhead{[K]} &
\colhead{[erg s$^{-1}$]} &
\colhead{[R$_{\odot}$]} &
\colhead{[mag]} &
\colhead{[M$_{\odot}$]} &
\colhead{Age} &
\colhead{Age} &
\colhead{Reg.} &
\colhead{Type} & 
\colhead{Class.} &
\colhead{Name} &
\colhead{ID} &
\colhead{Class.} &
\colhead{ID} &
\colhead{ID} 
\\
\colhead{} &
\colhead{[deg]} &
\colhead{[deg]} &
\colhead{[$^{\prime \prime}$]} &
\colhead{} &
\colhead{[$^{\prime \prime}$]} &
\colhead{[erg s$^{-1}$]} &
\colhead{} &
\colhead{} &
\colhead{} &
\colhead{} &
\colhead{} &
\colhead{[Myr]} &
\colhead{[Myr]} &
\colhead{ID} &
\colhead{} & 
\colhead{} &
\colhead{} &
\colhead{} &
\colhead{} &
\colhead{} &
\colhead{} 
}
\startdata
\hline \\ 
\multirow{2}{*}{192} & \multirow{2}{*}{23.375792} & \multirow{2}{*}{30.715470} & \multirow{2}{*}{0.7} & 013330.2309+304255.014 & 0.9 & \multirow{2}{*}{7.1e+34} & 4.5$^{+0.1}_{-0.1}$ & 4.2$^{+0.1}_{-0.1}$ & 4.4$^{+0.8}_{-0.1}$ & 0.4$^{+0.3}_{-0.1}$ & 11$^{+2}_{-1}$ & 15$^{+4}_{-10}$ & \multirow{2}{*}{34$^{+3}_{-22}$} & \multirow{2}{*}{1244} & B-MS & \multirow{2}{*}{stellar} & \multirow{2}{*}{013330.19+304255.6} & \multirow{2}{*}{---} & \multirow{2}{*}{---} & \multirow{2}{*}{---} & \multirow{2}{*}{---}  \\
 & & & & 013330.1734+304254.931 & 0.8 &  & 4.2$^{+0.1}_{-0.1}$ & 2.8$^{+0.3}_{-0.1}$ & 3.4$^{+0.4}_{-0.1}$ & 0.2$^{+0.4}_{-0.1}$ & 5$^{+1}_{-1}$ & 39$^{+49}_{-12}$ &  & & B-MS &  & & & &  \\
\hline \\ 
200 & 23.380500 & 30.567280 & 0.5 & 013331.2921+303402.538 & 0.5 & 7.9e+34 & 4.3$^{+0.1}_{-0.1}$ & 3.4$^{+0.1}_{-0.2}$ & 4.7$^{+0.5}_{-0.3}$ & 0.4$^{+0.2}_{-0.2}$ & 7$^{+1}_{-1}$ & 39$^{+19}_{-11}$ & 64$^{+9}_{-42}$ & 399 & B-MS & --- & 013331.32+303402.2 & --- & --- & --- & ---  \\
\hline \\ 
204 & 23.382167 & 30.598940 & 0.5 & 013331.7587+303555.916 & 0.6 & 6.0e+34 & 4.3$^{+0.1}_{-0.1}$ & 3.5$^{+0.1}_{-0.1}$ & 4.7$^{+0.1}_{-0.2}$ & 0.4$^{+0.1}_{-0.1}$ & 7$^{+1}_{-1}$ & 31$^{+8}_{-13}$ & 38$^{+18}_{-13}$ & 583 & B-MS & --- & --- & --- & --- & --- & ---  \\
\hline \\
\enddata
\end{deluxetable*}
\end{longrotatetable}

\begin{deluxetable*}{p{0.35\linewidth} p{0.6\linewidth}}
\tablecaption{Color and magnitude criteria used to select HMXB companion star candidates\label{table:comp_star_criteria}}
\tablehead{
\colhead{Type of star} &  
\colhead{Criteria}
}
\startdata
M$\gtrsim$8 M$_{\odot}$ MS star with 0$-$1 mag. A$_{V}$ &  [(F814W$<$24 and F475W$-$F814W$<$0.75) or (F814W$<$19)] and [(F336W$<$24 and F275W$-$F336W$<$1) or (F336W$<$22)]\\
M$\gtrsim$8 M$_{\odot}$ MS star with $\leq$2 mag. A$_{V}$ & [(F814W$<$24 and F475W$-$F814W$<$1.25)] and [(F336W$<$24 and F275W$-$F336W$<$1) or (F336W$<$22)] \\
M$\gtrsim$8 M$_{\odot}$ G/SG star with 0$-$1 mag. A$_{V}$ & [(F814W$<$23 and F475W$-$F814W$<$0.75) or (F814W$<$21)] and [F336W$<$24] \\
M$\gtrsim$8 M$_{\odot}$ G/SG star with $\leq$2 mag. A$_{V}$ & [(F814W$<$24 and F475W$-$F814W$<$1.5) or (F814W$<$22)] and [F336W$<$24]
\enddata
\tablecomments{Color and magnitude cuts used to select massive main sequence (MS) and giant/supergiant (G/SG) stars within the 1.5$\sigma$ positional errors of ChASeM33 X-ray source positions. We plot the stars selected using these criteria on a CMD in Figure \ref{fig:HMXB_CMDs}.}
\end{deluxetable*}

\newpage
\startlongtable
\setlength{\tabcolsep}{3pt}
\begin{deluxetable*}{cccccccc}
\tabletypesize{\scriptsize}
\tablecaption{PHATTER Photometry for HMXB Companion Star Candidates \label{table:phatter_photometry}}
\tablehead{
\colhead{PHATTER} &
\colhead{ChASeM33} &
\colhead{F110W} & 
\colhead{F160W} & 
\colhead{F814W} & 
\colhead{F475W} & 
\colhead{F336W} & 
\colhead{F275W}
\\
\colhead{Name} &
\colhead{ID} &
\colhead{[mag]} &
\colhead{[mag]} &
\colhead{[mag]} &
\colhead{[mag]} &
\colhead{[mag]} &
\colhead{[mag]}
}
\startdata
013330.2309+304255.014 & 192 & --- & --- & 22.50 & 22.44 & 21.12 & 20.83 \\ 
013330.1734+304254.931 & 192 & --- & --- & 23.97 & 23.96 & 23.09 & 23.12 \\ 
013331.2921+303402.538 & 200 & 22.88 & 22.30 & 23.18 & 23.14 & 22.10 & 21.71 \\ 
013331.7587+303555.916 & 204 & 23.22 & 23.02 & 23.11 & 23.05 & 22.04 & 21.98 \\ 
013331.8256+304011.190 & 205 & 23.73 & 23.70 & 23.72 & 23.72 & 22.89 & 22.83 \\ 
013331.8551+304012.079 & 205 & 23.03 & 22.70 & 23.73 & 23.94 & 23.88 & 24.81 \\ 
013332.1735+303656.933 & 207 & 19.60 & 18.85 & 20.12 & 21.17 & 20.75 & 21.06 \\ 
013332.5421+303618.159 & 212 & 23.41 & 23.48 & 23.97 & 24.02 & 23.93 & 25.17 \\
\enddata
\tablecomments{Multi-band photometry for each identified HMXB companion star candidate. We list the name of each star (derived from its position) in the PHATTER catalog and the ChASeM33 ID number of the X-ray source each HMXB companion star candidate is associated with for ease in cross-referencing Table \ref{table:beast_table}. Photometry comes from the full PHATTER photometric catalog presented in \citet{Williams2021}.}
\end{deluxetable*}

\begin{deluxetable}{cccc}
\tablecaption{Physical Properties Used to Classify Companion Stars\label{table:sp_type_ranges}}
\tablehead{
\colhead{Stellar Type} &  
\colhead{$T_{eff}$ [K]} & 
\colhead{log(L) [$L_{\odot}$]} & 
\colhead{Radius [$R_{\odot}$]}
}
\startdata
O-MS & 30,000$-$44,500 & 4.72$-$5.90 & 8.4$-$15 \\
B-MS & 9,250$-$30,000 & 1.73$-$4.72 & 2.7$-$8.4 \\
O-G & 29,000$-$42,500 & 5.04$-$6.00 & 13$-$18 \\
B-G & 10,100$-$29,000 & 2.02$-$5.04 & 3.4$-$13 \\
O-SG  & 26,000$-$40,300 & 5.52$-$6.04 & 22$-$25 \\
B-SG & 9730$-$26,000 & 4.54$-$5.52 & 25$-$66 \\
\enddata
\tablecomments{List of the ranges of effective temperatures, luminosities, and radii used to classify companion stars. We plot these ranges as colored patches on the Hertzsprung-Russell diagram in Figure \ref{fig:HR_diagram}. Values come from Appendix B of \citet{Lamers2017}.}
\end{deluxetable}

\begin{longrotatetable}
\setlength{\tabcolsep}{3pt}
\begin{deluxetable*}{ccccccccc}
\tabletypesize{\scriptsize}
\tablecaption{Values used to determine best HMXB candidate sample \label{table:summary_table_values}}
\tabletypesize{\scriptsize}
\tablehead{
\colhead{T11 ID} &
\colhead{PHATTER Pos.} &
\colhead{Chandra} &
\colhead{Chandra} &
\colhead{F110W$-$F160W} &
\colhead{SFR $<$ 80 Myr} &
\colhead{N$_{H}$} & 
\colhead{SED-derived A$_{V}$} &
\colhead{SFH-derived A$_{V}$}
\\
\colhead{} &
\colhead{} &
\colhead{HR1} &
\colhead{HR2} &
\colhead{[mag]} &
\colhead{[$\times10^{-4}$ M$_{\odot}$ yr$^{-1}$]} &
\colhead{[$\times10^{21}$ cm$^{-2}$]} &
\colhead{[mag]} &
\colhead{[mag]}
}
\startdata
192 & 013330.2309+304255.014 & 0.3 & -0.3 & -- & 0.2 & -- & 0.4$^{+0.29}_{-0.05}$ & 0.21\\
192 & 013330.1734+304254.931 & 0.3 & -0.3 & -- & 0.2 & -- & 0.2$^{+0.37}_{-0.04}$ & 0.21\\
200 & 013331.2921+303402.538 & 0.6 & 0.3 & 0.6 & 1.4 & -- & 0.4$^{+0.18}_{-0.24}$ & 0.11\\
204 & 013331.7587+303555.916 & -0.0 & 0.2 & 0.2 & 2.8 & -- & 0.4$^{+0.15}_{-0.14}$ & 0.21\\
205 & 013331.8256+304011.190 & -0.2 & 0.2 & 0.0 & 0.0 & -- & 0.2$^{+0.39}_{-0.08}$ & 0.11\\
205 & 013331.8551+304012.079 & -0.2 & 0.2 & 0.3 & 0.0 & -- & 0.8$^{+0.3}_{-0.1}$ & 0.11\\
207 & 013332.1735+303656.933 & 0.1 & 0.7 & 0.7 & 3.8 & 99.1$^{+9.91}_{-9.91}$ & 2.2$^{+0.14}_{-0.14}$ & 0.21\\
212 & 013332.5421+303618.159 & -0.1 & -0.1 & -0.1 & 2.6 & -- & 0.6$^{+0.39}_{-0.01}$ & 0.21\\
\enddata
\tablecomments{We present the multi-wavelength properties used to evaluate the quality of each HMXB candidate. For a full discussion of how we determined the highest quality HMXB candidate sample, see Section \ref{sec:flags}. The Chandra hardness ratios listed here were calculated using count rates from \citet{Tullmann2011}. The F110W-F160W color is calculated using the HST photometry for each HMXB companion star candidate, listed in Table \ref{table:phatter_photometry}. We include the mean SFR over the last 80 Myr, calculated from the recent SFH maps of M33 generated with the PHATTER optical photometry, presented in \citet{Lazzarini2022}. Spectral fits were performed on X-ray sources in the ChASeM33 catalog with sufficient counts, and we present the resulting best-fit N$_{H}$ for these sources from \citet{Tullmann2011}. The SED-derived A$_{V}$ value for each HMXB companion star candidate comes from our SED fits, and is also listed in Table \ref{table:beast_table}. The SFH-derived A$_{V}$ value comes from \citet{Lazzarini2022}.}
\end{deluxetable*}
\end{longrotatetable}

\begin{longrotatetable}
\setlength{\tabcolsep}{3pt}
\begin{deluxetable*}{ccccccccc}
\tabletypesize{\scriptsize}
\tablecaption{Flags used to determine best HMXB candidate sample \label{table:summary_table_flags}}
\tablehead{
\colhead{T11 ID} &
\colhead{PHATTER Pos.} &
\colhead{FLAG: Chandra} &
\colhead{FLAG: FG} &
\colhead{FLAG: Flat} &
\colhead{FLAG: No} &
\colhead{FLAG: N$_{H}$} &
\colhead{FLAG: Spectral} &
\colhead{FLAG} \\
\colhead{} &
\colhead{} &
\colhead{HRs} &
\colhead{IR Color} &
\colhead{SED shape} &
\colhead{SF $<$ 80 Myr} &
\colhead{A$_{V}$ mismatch} &
\colhead{Type} &
\colhead{SUM}
}
\startdata
192 & 013330.2309+304255.014 & 0 & 0 & 0 & 0 & 0 & 0 & 0\\ 
192 & 013330.1734+304254.931 & 0 & 0 & 0 & 0 & 0 & 0 & 0\\ 
200 & 013331.2921+303402.538 & 0 & 1 & 0 & 0 & 0 & 0 & 1\\ 
204 & 013331.7587+303555.916 & 0 & 0 & 0 & 0 & 0 & 0 & 0\\ 
205 & 013331.8256+304011.190 & 0 & 0 & 0 & 1 & 0 & 0 & 1\\ 
205 & 013331.8551+304012.079 & 0 & 0 & 0 & 1 & 0 & 0 & 1\\ 
207 & 013332.1735+303656.933 & 0 & 1 & 1 & 0 & 1 & 0 & 3\\ 
212 & 013332.5421+303618.159 & 0 & 0 & 0 & 0 & 0 & 0 & 0\\
\enddata
\tablecomments{For each identified HMXB companion star candidate, we list whether it raised each of our flags used to evaluate the quality of HMXB candidates. The values used to determine the flags are listed in Table \ref{table:summary_table_values} and a discussion of criterion is in Section \ref{sec:flags}.}
\end{deluxetable*}
\end{longrotatetable}

\begin{deluxetable}{ccc}
\setlength{\tabcolsep}{1.5pt}
\tablecaption{M33 HMXB Production Rates per Unit SFR\label{table:HMXB_prod_rates}}
\tablehead{
\colhead{ } &  
\colhead{HMXB Prod. Rate} & 
\colhead{HMXB Prod. Rate} \\
\colhead{Sample} &
\colhead{0$-$50 Myr ago}&
\colhead{0$-$80 Myr ago}\\
\colhead{} &
\colhead{[HMXBs/(M$_{\odot}$ yr$^{-1}$)]}&
\colhead{[HMXBs/(M$_{\odot}$ yr$^{-1}$)]}
}
\startdata
Full sample & 120$-$153 & 181$-$240\\
$\leq$1 Flag sample & 117$-$150  & 175$-$233 \\
0 Flag sample & 107$-$136 & 150$-$199
\enddata
\tablecomments{HMXB production rates calculated per unit SFR for our HMXB candidate sample. The rows correspond to subsamples based on our evaluation system described in more detail in Section \ref{sec:flags}. The samples used for the calculations in each row match the samples in Figure \ref{fig:age_prod_rate}.}
\end{deluxetable}

\begin{deluxetable*}{cccc}
\tablecaption{Mean SFR in M31 and M33 over different timescales\label{table:m31_m33_sfr}}
\tablehead{
\colhead{Galaxy} &  
\colhead{50 Myr mean SFR [M$_{\odot}$ yr$^{-1}$]} & 
\colhead{80 Myr mean SFR [M$_{\odot}$ yr$^{-1}$]} & 
\colhead{100 Myr mean SFR [M$_{\odot}$ yr$^{-1}$]}
}
\startdata
M31 & 0.23$\pm$0.01 & 0.24$\pm$0.01 & 0.24$\pm$0.01 \\
M33 & 0.29$\pm$0.04 & 0.31$\pm$0.04 & 0.32$\pm$0.02\\
\enddata
\tablecomments{Mean SFR over various timescales in M31 and M33 in the areas of each galaxy covered by overlapping HST and Chandra observations. The SFR values come from spatially resolved recent SFHs of each galaxy \citep{Lewis,Lazzarini2022}.}
\end{deluxetable*}


\end{document}